\pdfoutput=1
\documentclass[showpacs,nofootinbib,showkeys,eqsecnum,prd,aps,preprint]{revtex4-2}

\usepackage[english]{babel}

\usepackage{amssymb}
\usepackage{amsmath}
\usepackage{amsfonts}
\usepackage{latexsym}
\usepackage{mathrsfs}
\usepackage{bm}
\usepackage{tensor}
\usepackage{hyperref}

\usepackage{bm}
\usepackage{tensor}
\usepackage{hyperref}

\usepackage{graphicx} 
\usepackage{epsfig}
\usepackage{indentfirst,float,wrapfig}

\usepackage{xcolor}

\usepackage{graphicx}

\DeclareGraphicsExtensions{.pdf,.png,.jpg}

\begin{document}

\title{Quasiparticle solutions to the 1D  nonlocal Fisher--KPP equation with a fractal time derivative in the weak diffusion approximation
}

\author{A.~V.~Shapovalov$^{1,2}$}
\newcommand{\orcidauthorA}{0000-0003-2170-1503}
\email{shpv@mail.tsu.ru}
\author{S.~A.~Siniukov${}^{1}$}
\email{ssaykmh@yandex.ru}
\affiliation{${}^1$Department of Theoretical Physics, Tomsk State University, Novosobornaya Sq. 1, 634050 Tomsk, Russia}
\affiliation{${}^2$Laboratory for Theoretical Cosmology, International Centre of Gravity and Cosmos, Tomsk State University of Control Systems and Radioelectronics, 40 Lenina av., 634050 Tomsk, Russia}

\begin{abstract}
In this paper, we propose an approach for constructing quasiparticle-like asymptotic solutions within the weak diffusion approximation for the generalized population Fisher--Kolmogorov--Petrovskii--Piskunov (Fisher--KPP) equation, which incorporates nonlocal quadratic competitive losses and a fractal time derivative of non-integer order $\alpha$, where $0<\alpha\leq 1$. This approach is based on the semiclassical approximation and the principles of the Maslov method. The fractal time derivative is introduced in the framework of $F^{\alpha}$-calculus. The Fisher--KPP equation is decomposed into a system of nonlinear equations that describe the dynamics of interacting quasiparticles within classes of trajectory-concentrated functions. A key element in constructing approximate quasiparticle solutions is the interplay between the dynamical system of quasiparticle moments and an auxiliary linear system of equations, which is coupled with the nonlinear system. General constructions are illustrated through examples that examine the effect of the fractal parameter $\alpha$ on quasiparticle behavior.
\end{abstract}



%
%


\maketitle

\section{Introduction}
\label{int}

The modeling of reaction-diffusion (RD) systems serves as a fundamental theoretical framework for studying nonlinear phenomena across diverse fields, including physics, chemistry, biology, and engineering \cite{murray2002}. To accurately describe RD processes in complex physical and biological systems exhibiting anomalous diffusion and memory effects, a fractal derivative with respect to time is often introduced in the mathematical formulation.
\par
Over the past few decades, the development of fractal calculus has enabled the formulation of model equations for physical and biological processes and systems directly in terms of fractal operators, such as derivatives and integrals, much like in ordinary calculus.
The basic concepts of fractal calculus are outlined, for instance, in \cite{prgn2009,prgn2011,prsatgn2011,goltunnigol2019,golman2019,golman2022} and the references therein.
These advancements have sparked  a growing interest in research into physical phenomena on fractal objects, as partially demonstrated in \cite{golman2019,golman2022}. In particular, fractal calculus provides an adequate framework for simulating the dynamics of reaction-diffusion (RD) systems in media with complex properties and irregular influences.
\par
The emergence of model equations incorporating fractal operators poses significant challenges for developing methods to analyze these equations and to obtain both exact and approximate solutions.
\par
In the study of biological populations comprising a single species, the classical Fisher--Kolmogorov--Petrovskii--Piskunov (Fisher--KPP) equation of the RD type that accounts for local competitive losses was introduced in \cite{fisher1937,kpp1937} to describe population waves. Incorporating nonlocal competitive losses, which enable long-range interactions within the population \cite{murray2002}, expands the range of dynamic regimes that can be modeled.
The modified Fisher--KPP equation with nonlocal competitive losses provides a framework for describing the formation and evolution of patterns in single-species populations. Similar nonlocal kinetic models have been primarily studied through numerical simulations (e.g., \cite{murray2002,lopez2004,lopez2014,lima2022}). Analytical methods have been developed under appropriate approximations.
The formalism for constructing asymptotic solutions within the weak diffusion approximation was developed in \cite{trsh2009,shtr2018,shtr2019} for the nonlocal Fisher--KPP equation, which includes the standard first-order time derivative:
\begin{align}
&- u_t(x,t)+\varepsilon u_{xx}(x,t)+a(x,t) u(x,t)-\varkappa u(x,t)\int_{-\infty}^{\infty}b(x,y,t)u(y,t)d y=0.
\label{fkpp}
\end{align}
Here the equation is written in dimensionless form, a real smooth function $u(x,t)$ is a population density, $u(x,t)\to 0$ as $|x|\to \infty$; $u_t=\partial_t u=\partial u/\partial t$, $u_x=\partial_x u=\partial u/\partial x$, $u_{xx}=\partial^2 u/\partial x^2$;  $a(x,t)$ and  $b(x,y,t)$ are given infinitely smooth functions increasing, as $|x|\,,|y|\to\infty$, no faster than polynomially; $\varkappa \,(>0)$ is a real nonlinearity parameter. The term $-\varkappa u(x,t)\int\limits^\infty_{-\infty}b(x,y,t) u(y,t)dy$ stands for the nonlocal competition losses and is characterized by an influence function $b(x,y,t)$, the coefficient $a(x,t)$ stands for the reproduction rate, and $\varepsilon$ is the diffusion coefficient.
In the case of local competitive losses, the integral term in equation \eqref{fkpp} is replaced by $-\varkappa b u^2$.
\par
\par
In \cite{kulsh2024}, the approach developed in \cite{trsh2009, shtr2018}, which utilizes the Maslov  method (see, e.g., \cite{maslov1994, beldob1992}), was generalized to construct asymptotic solutions for equation \eqref{fkpp} in the form of two quasiparticles. These solutions characterize the dynamics of interacting local modes of population density.
\par
In this paper, we extend the formalism developed in \cite{trsh2009, shtr2018, shtr2019}, and \cite{kulsh2024} to the nonlocal Fisher--KPP equation with a fractal time derivative of fractional order $\alpha$. Within the weak diffusion approximation, we derive asymptotic solutions that capture the fractal dynamics of quasiparticles. Preliminary results were presented in \cite{shsin2024}.

Note that the concepts and techniques of the Maslov method, originally developed in quantum theory, have found broad applications in a range of linear and nonlinear mathematical physics problems. For further details, refer to \cite{maslov1994, beldob1992}, and \cite{kulsh2024}, along with the references therein. However, this approach has not been previously applied to problems involving fractal dynamics.
\par
Fractional powers in dimensional indices emerge when describing fractal media. In such media, unlike in continuous ones, a randomly walking particle can  move away from its starting point more slowly because not all directions of motion are accessible, and only those aligned with the fractal structure (e.g., on a plane or in space). The slowdown of diffusion in fractal media may manifest itself in such a way that physical quantities evolve more slowly than the first space derivative would suggest. This effect can be captured either through an integro-differential equation involving a fractional-order space derivative or, more logically, by directly employing a fractal derivative defined explicitly on the fractal structure.
In the case of temporal fractality, the random motion of a particle occurs in a certain sense in a jumpy manner at moments  associated with the fractal time interval. This behavior might be expected to accelerate diffusion.
\par
The properties of fractional derivatives, which reflect the ``memory'' characteristics of a process, and fractal derivatives, which characterize local self-similarity, have been extensively discussed and applied to RD type equations, including the Fisher--KPP equation, in numerous publications such as \cite{hilant1995, klwgang1998, barl1998, yzmbl2016}, as well as in recent works, for example, \cite{golman2022, ishii2024, awon2025}.
The inclusion of a fractal time derivative in the Fisher--KPP equation enables the incorporation of features specific to population dynamics when fractal properties arise from either the population habitat or the structural organization of the population itself.
We investigate the Fisher--KPP equation with a fractal time derivative within the framework of the $F^{\alpha}$-calculus \cite{prgn2009,prgn2011,prsatgn2011,goltunnigol2019,golman2019}.
\par
The structure of this paper is as follows: Section \ref{sect1} provides
a brief overview of the necessary basic concepts and notations from the fractal calculus and introduces the nonlocal Fisher--KPP equation with a fractal time derivative.
\par
In Section \ref{sec-tcf1}, we introduce the class of trajectory-concentrated functions in which asymptotic solutions are sought. A definition of quasiparticles is provided, and the Fisher--KPP equation under consideration is decomposed into a system of equations describing the fractal dynamics of interacting quasiparticles.
\par
In Section \ref{sect-estim}, we derive estimates for operators in the class of trajectory-concentrated functions and introduce the moments that characterize quasiparticles. The expansions of the coefficients of the operators in the equation are performed within this function class.
\par
In Section \ref{sect-momdyn}, we derive a fractal dynamical system for the moments of quasiparticles and discuss its properties. Using the general solution of the dynamical system of moments, in Section \ref{sect-ale}, we introduce a system of linear equations with a fractal time derivative, associated with the equations for quasiparticles.
\par
In Section \ref{sect-cauch}, we explore the relationship between the Cauchy problem for the fractal nonlocal Fisher--KPP equation and the associated linear system under certain conditions.
In the asymptotic approximation, we derive the Green function for the associated linear system, which generates approximate solutions to the original nonlinear system describing the quasiparticles.
\par
In Section \ref{exam-sol1}, we illustrate the general results with examples of the dynamics of two quasiparticles and examine the influence of the fractal derivative parameter on quasiparticle dynamics.
\par
In Section \ref{sect-conc}, we provide our concluding remarks.

\section{Fisher--KPP equation with fractal time derivative}
\label{sect1}

In equation \eqref{fkpp}, we assume that time $t$ does not evolve continuously, but rather on a fractal set $F \subset I$, where $I = I[a, b]$ (with $a < b$) is a closed interval of the real line $\mathbb{R}^1$. In this context, the time derivative $u_t(x, t)$ in \eqref{fkpp} must be replaced by an appropriate time derivative on $F$. To accomplish this, we employ the $F^\alpha$ calculus, as developed in \cite{prgn2009, prgn2011}, with additional insights from \cite{golman2022}.
For the fractal set $F$ we  consider a $C^{\zeta}$ Cantor set (see, e.g., \cite{prgn2009,prgn2011,golman2022,glfrglb2018,shbr2021}).
Here we provide a brief overview of the $F^\alpha$ calculus,  primarily following \cite{prgn2009,prgn2011,golman2022,shbr2021}.
\par
The {\it indicator (flag) function} of the set $F$ and $I[a,b]$
is given as
\begin{align}
&\theta(F,I)=\left\{
\begin{array}{cc}
1 ,& {if}\,\, F\cap I \neq  \emptyset ,\\
 0 ,&  {otherwise} .
\end{array}
\right.
\label{flg}
\end{align}
Let $P_{[a,b]}=\{a=x_0, x_1, \ldots , x_n=b \}$, $x_i < x_{i+1}$, $i=0,1,\ldots ,n-1$, $|P|=\max\limits_{0\leq i\leq n-1}(x_{i+1}-x_i)\leq\delta$, $\delta\geq 0$, be a $\delta$-partition of the interval $I$.
\par
For a parameter $0<\alpha\leq 1$ and $\delta$-partitions $P_{[a,b]}$ of $I$, {\it the coarse-grained mass function} $\gamma^\alpha_{\delta}(F,a,b)$  is given by the expression
\begin{align}
&\gamma^{\alpha}_{\delta}(F,a,b)=\inf\limits_{\{P_{[a,b]}:|P|\leq\delta\}}
\sum_{i=0}^{n-1}\Gamma^{-1}(\alpha+1)(x_{i+1}-x_{i})^\alpha \theta(F,[x_i,x_{i+1}]),
\label{cgmss-1}
\end{align}
and {\it the Haussdorff mass measure} is defined as
\begin{align}
&\gamma^{\alpha}(F,a,b)=\lim\limits_{\delta\to 0}\gamma^{\alpha}_{\delta}(F,a,b).
\label{mssm-1}
\end{align}
The infimum in \eqref{cgmss-1} is taken over all $\delta$-partitions $P_{[a,b]}$ of the interval $I$, and $\Gamma(x)$ is Euler's Gamma function.
Then {\it the fractal ($\gamma$-)dimension} of the set $F$ (the value of $\alpha$ at which $\gamma^\alpha$ is finite) is defined by the condition
\begin{align}
\dim_{\gamma}(F\cap [a,b])=&\inf\{\alpha : \gamma^{\alpha}(F,a,b)=0\},\cr
 =&\sup\{\alpha : \gamma^{\alpha}(F,a,b)=+\infty\}.
\label{dim-1}
\end{align}
For a fixed number $a_0\in \mathbb{R}^1$, {\it the integral staircase function} $S^{\alpha}_{F}(x)$ on $F$ is defined as
\begin{align}
&S^{\alpha}_{F}(x)=\left\{
\begin{array}{cc}
\gamma^\alpha(F, a_0,x),& {if}\,\, x\geq a_0,\\
- \gamma^\alpha(F,x, a_0),&{otherwise} .
\end{array}
\right.
\label{stcf-1}
\end{align}
The function $S^{\alpha}_{F}(x)$ increases monotonically with $x$, $S^{\alpha}_{F}(y)-S^{\alpha}_{F}(x)=\gamma^\alpha(F,x,y)$, $S^{\alpha}_{F}$ is continuous on $(a,b)$, and $S^{\alpha}_{F}$ is
a constant in $[x,y]$ when $F\cap (x,y)=\emptyset$.
\par
For a function $f:\mathbb{R}^1\rightarrow\mathbb{R}^1$, $F\subset\mathbb{R}^1$ and $x\in F$, the number $l$ is termed {\it $F$-limit} of the function $f$ through the set $F$ as $y\to x$, $y\in F$, if $\forall \varepsilon >0$, there exists $\delta >0$ such that $|f(y)-l|<\varepsilon$ when $|y-x|<\delta$ \cite{prgn2009}. The following notation is used: $l=F-\lim_{y\to x} f(y)$.
\par
The function $f$ is {\it F-continuous} at the point $x\in F$, if $f(x)=F-\lim_{y\to x} f(y)$. Note that $F$-continuity is not defined for $x\notin F$.
\par
Using the integral staircase function $S^{\alpha}_{F}(x)$ for the set $F$, we can introduce  the important concept of $\alpha$-perfect set. This set can be constructed algorithmically for $F$ and represents its essential fractal properties. For $\alpha$-perfect set $F$ the following property holds: if $x\in F$, then $S^\alpha_F(y)$ is different from $S^\alpha_F(x)$ at all points y on at least one side of $x$ (see \cite{prgn2009} for details). This property makes it possible to define the derivative on a fractal  set.
An example of an $\alpha$-perfect set is the middle $C^{\zeta}$ Cantor set for $\alpha=\zeta$ where $\zeta$ is the Haussdorff dimension of $C^{\zeta}$, in particular, for the $1/3$-middle Cantor  set $\dim C^{\zeta}=\log 2/\log 3$  (e.g., \cite{golman2022}).
\par
For a perfect set $F$ and a function $f$, the {\it $F^\alpha$-derivative} of $f$ at $x$ is defined as \cite{prgn2009}
\begin{align}
&D^{\alpha}_{F} f(x)=\left\{
\begin{array}{cc}
F-\lim\limits_{y\to x}\frac{f(y)-f(x)}{S^\alpha_F(y)-S^\alpha_F(x)},& {if}\,\, x\in F,\\
0,&{otherwise},
\end{array}
\right.
\label{fdir-1}
\end{align}
when the $F$-limit exists.
\par
Some basic ruiles for $F^\alpha$-differentiation are given below following \cite{prgn2009}.
\par
The derivative of a constant fuinction is zero: if $f(x)=c$, $c$ is a real constant, $x\in[a,b]$, then
$D^{\alpha}_{F} c=0$.
\par
The derivative of the integral staircase function $S^\alpha_F(x)$, $x\in[a,b]$, is the
characteristic function $\chi_F(x)$ of $F$:
\begin{align}
D^{\alpha}_{F} S^\alpha_F(x)= \chi_F(x),\quad
\chi_F(x)=\left\{
\begin{array}{cc}
1, & \,\, x\in F,\\
0,& \,\, x\notin F.
\end{array}
\right.
\label{dchf-1}
\end{align}
Let $f$ and $g$ be functions defined on $[a,b]$, and suppose the derivatives
$D^{\alpha}_{F} f(x)$ and $D^{\alpha}_{F} g(x)$ exist for all $x\in[a,b]$, then the following properties hold.
\par
Linearity: The  derivatives  $D^{\alpha}_{F} (c f(x))$ and
$D^{\alpha}_{F} (f(x)+g(x))$ exist, where $c$ is a constant. Moreover,  $D^{\alpha}_{F}(c f(x))=c D^{\alpha}_{F} f(x)$  and $D^{\alpha}_{F}(f(x)+g(x))=D^{\alpha}_{F} f(x)+D^{\alpha}_{F} g(x)$.
\par
Leibniz rule: The derivative $D^\alpha_F(f(x)g(x))$ exists and satisfies
\begin{align}
&D^\alpha_F(f(x)g(x))=D^\alpha_F(f(x))g(x)+f(x)D^\alpha_Fg(x).
\label{leib-1}
\end{align}
\par
Chain rule: The chain rule for the fractal derivative is given by (see \cite{ashgol2018}):
\begin{align}
&D^\alpha_F\big(f\big( g(x)\big)\big)=D^\alpha_F f\big(g(x)\big)\, D^\alpha_Fg(x).
\label{chrl-1}
\end{align}
\par
In particular, $D^\alpha_F\big(f(x)/g(x)\big)=g^{-2}(x)(g(x)D^\alpha_Ff(x)-f(x)D^\alpha_Fg(x))$.
\par
\par
Let $B(F)$ denote the class of bounded functions on $F$.
Next, to  introduce the $F^\alpha$-integral of a  function $f(x)\in B(F)$ on $F\subset I$, the following definitions are used (see, e.g., \cite{prgn2009,golman2022}):
\begin{align}
&M[f,F,I]=\left\{
\begin{array}{cc}
\sup\limits_{x\in F\cap I} f(x), & \,\, x\in F,\\
0,& \,\, otherwise,
\end{array}
\right.\,\,
m[f,F,I]=\left\{
\begin{array}{cc}
\inf\limits_{x\in F\cap I} f(x), & \,\, x\in F,\\
0,& \,\, otherwise .
\end{array}
\right.
\label{int-1}
\end{align}
The upper $F^\alpha$-sum ($U^\alpha[f,F,P]$)
 and the lower $F^\alpha$-sum ($L^\alpha[f,F,P]$) over the subdivision $P=\{x_0,x_1,\ldots ,x_n\}$  of the interval $I$ for a function $f\in B(F)$ and the finite function $S^\alpha_F(x)$, $x\in I$, are given by the following expressions:
 \begin{align}
 &U^\alpha[f,F,P]=\sum\limits_{i=0}^{n-1}M[f,F,[x_i,x_{i+1}]]\big( S^\alpha_F(x_{i+1})-S^\alpha_F(x_{i})\big),\cr
 &L^\alpha[f,F,P]=\sum\limits_{i=0}^{n-1}m[f,F,[x_i,x_{i+1}]]\big( S^\alpha_F(x_{i+1})-S^\alpha_F(x_{i})\big).
  \label{int-1a}
 \end{align}
 Then the $F^\alpha$-integral of a function $f(x)\in B(F)$ is defines as
 \begin{align}
 &\int\limits_{a}^b f(x) d^\alpha_F x=
 \sup\limits_{P_{[a,b]}} L^\alpha [f,F,P]=
 \inf\limits_{P_{[a,b]}} U^\alpha [f,F,P].
  \label{int-2a}
 \end{align}
If the integral \eqref{int-2a} exists, the function $f(x)\in B(F)$ is called $F^\alpha$-integrable on $I=[a,b]$.
\par
The $F^\alpha$-integral has the obvious property of linearity, and also
$\int_a^b f(x)d^\alpha_F x=$ $-\int_b^a f(x)d^\alpha_F x$.
Given below are some basic properties that characterize the specifics of the $F^\alpha$-integral (see \cite{prgn2009} for details).
\par
For a continous and $F^\alpha$-differentiable function $f:\mathbb{R}^1\rightarrow\mathbb{R}^1$, $\alpha$-perfect set $F$ and $F$-continuous function $h:\mathbb{R}^1\rightarrow\mathbb{R}^1$ such that
 $h(x)\chi_F(x)=D^\alpha_F f(x)$, the following
 equality holds:
 \begin{align}
 \int_a^b h(x) d^\alpha_F x=f(b)-f(a).
 \label{int-2aa}
\end{align}
Also, we have
 \begin{align}
 \int_a^b \chi_F(x) d^\alpha_F x=S^\alpha_F(b)-S^\alpha_F(a).
 \label{int-2ab}
\end{align}
The formula for integration by parts reads
 \begin{align}
 \int_a^b f(x) g(x)d^\alpha_F x=
 \left[f(x)\int_a^x g(y)d^\alpha_F y\right]\bigg{|}_a^b-
 \int_a^b \big(D^\alpha_F f(x)\big)
 \int_a^x g(y)d^\alpha_F y \,d^\alpha_F x.
 \label{int-2ac}
\end{align}
We now express the nonlocal Fisher--KPP equation \eqref{fkpp} with the fractal time derivative for the function $u(x,t)$ in the following form:
\begin{align}
&- D^\alpha_{F,t}u(x,t)+\chi_F(t)\left(\varepsilon u_{xx}(x,t)+a(x,t) u(x,t)-\varkappa u(x,t)\int_{-\infty}^{\infty}b(x,y,t)u(y,t)d y\right)=0,
\label{flfkpp1}
\end{align}
where $D^\alpha_{F,t}$ denotes the partial $F^\alpha$-derivative
given by \eqref{fdir-1} with respect to $t$, $x\in \mathbb{R}^1$,  $F\subset [0,T]\subset \mathbb{R}^1$, and $t\in [0,T]$.

\section{Trajectory concentrated functions and decomposition of the equation}
\label{sec-tcf1}

We investigate the influence of time fractality on the dynamics described by equation \eqref{flfkpp1}. Specifically, we analyze the asymptotic solutions of equation \eqref{flfkpp1} under the weak diffusion approximation, considering the parameter $\varepsilon$ as a small asymptotic parameter ($\varepsilon\to 0$).
\par
Let $X(\varepsilon)$  and $R(\varepsilon)$ be $F^{\alpha}$- continuous and $F^{\alpha}$-differentiable real functions of $t$ in the sense of $F^\alpha$-calculus (Section \ref{sect1}) and regularly depend on $\sqrt{\varepsilon}$ as $\varepsilon\to 0$,
$X(\varepsilon):F\rightarrow \mathbb{R}^1$, $R(\varepsilon):F\rightarrow \mathbb{R}^1$.
The dependence on $t$ in the functions $X(t,\varepsilon)$ and $R(t,\varepsilon)$ is mediated by the integral staircase function $S_F^\alpha(t)$.
\par
Asymptotic solutions for equation \eqref{flfkpp1}, as well as for equation \eqref{fkpp}, can be constructed in the limit as $\varepsilon\to 0$ within the framework of the so-called trajectory concentrated functions (TCF) \cite{shtr2018,shtr2019}. The class $\mathcal{P}_{(\alpha)t}^{\varepsilon}$ of these functions is defined by its common element as follows:
\begin{equation}
\label{eq1}
\mathcal{P}_{(\alpha)t}^{\varepsilon}=\mathcal{P}_{(\alpha)t}^{\varepsilon} \big(X(\varepsilon),R(\varepsilon) \big)=
  \left\{ \Phi :\Phi(x,t,\varepsilon)=\frac{1}{\sqrt{\varepsilon}}
   \exp\left(\frac{1}{\varepsilon}R(t,\varepsilon)\right)\varphi\left( \frac{\Delta x}{\sqrt \varepsilon},t,\varepsilon\right)
\right\},
\end{equation}
where  $\Delta x=x-X(t,\varepsilon)$, $x\in\mathbb{R}^1$; the real functions  $X(t,\varepsilon)$, $R(t,\varepsilon)$ are functional parameters of the class $\mathcal{P}_{(\alpha)t}^{\varepsilon}$, and $\varphi\left( \xi,t,\varepsilon\right)$
  regularly depend on $\sqrt{\varepsilon}$ as $\varepsilon\to 0$, and  $\varphi$
  belongs to the Schwartz space $\mathbb{S}$ with respect to the argument $\xi$.
\par
The functions of the class  $\mathcal{P}_{(\alpha)t}^{\varepsilon}$ are localized within a neighborhood of a point that moves in the phase space of a dynamical system associated with the moments of the equation solution. Furthermore, $X(t,\varepsilon)$ characterizes the spatial trajectory of this point (for details, see \cite{trsh2009,shtr2018}).
\par
We also note that the class of trajectory-concentrated functions has previously been used in quantum mechanics (see \cite{bgbltr1996} and references therein). In this context, approximate solutions of quantum equations constructed within this class are interpreted as semiclassical solutions. By analogy, we adopt this terminology for the Fisher--KPP equations \eqref{fkpp} and \eqref{flfkpp1}.
 \par
To construct quasiparticle solutions to equation \eqref{flfkpp1}, we employ a collection of the classes
\begin{align}
&\mathcal{P}_{(\alpha)t,s}^{\varepsilon}=\mathcal{P}_{(\alpha)t}^{\varepsilon} \big(X_s(\varepsilon),R_s(\varepsilon) \big)
 \label{eq1a}
 \end{align}
 with functional parameters $X_s(t,\varepsilon)$ and $R_s(t,\varepsilon)$. Here $s=1,2, \ldots , K$, and $K$ denotes the number of quasiparticles.
 The function $X_s(t,\varepsilon)$ corresponds to the trajectory of $s$-th quasiparticle.

We seek a solution to equation \eqref{flfkpp1} in the form
\begin{align}
&u(x,t,\varepsilon)=\sum_{s=1}^{K} u_{s}(x,t,\varepsilon),
\label{sol1a}
\end{align}
where the functions $u_{s}(x,t,\varepsilon)\in \mathcal{P}_{(\alpha)t,s}^{\varepsilon}$
are governed by the equations
\begin{align}
&- D^\alpha_{F,t}u_s(x,t,\varepsilon)+ \chi_F(t)\Big(\varepsilon u_{s\,xx}(x,t,\varepsilon)+a(x,t) u_s(x,t,\varepsilon)
\cr
&-\varkappa u_s(x,t,\varepsilon)\int_{-\infty}^{\infty}b(x,y,t)\sum_{\bar{s}=1}^{K}u_{\bar{s}}(y,t,\varepsilon)d y\Big)=0.
\label{flsys1a}
\end{align}
The summation of  equations \eqref{flsys1a} yields equation \eqref{flfkpp1} for the function $u(x,t,\varepsilon)$ defined in  \eqref{sol1a}. Notably, the function  $u(x,t,\varepsilon)$ in \eqref{sol1a} cannot be regarded as a superposition of the functions  $u_s(x,t,\varepsilon)$ , since these functions are interdependent. Here, $u_s$ represents the  $s$-th quasiparticle. Accordingly, we refer to equations \eqref{flsys1a} as the quasiparticle decomposition system (QDS) for $K$ quasiparticles.
\par
Next, for  asymptotic solution  constructing to equation \eqref{flfkpp1}, we will define estimates of corresponding operators in the classes $\mathcal{P}_{(\alpha)t,s}^{\varepsilon}$.

\section{Estimates of operators and moments}
\label{sect-estim}

In accordance with \cite{trsh2009,shtr2018} and \cite{kulsh2024}, an  operator $\hat{A}$, acting on functions $u_s$ from the class $\mathcal{P}_{(\alpha)t,s}^{\varepsilon}$, has the  estimate $\hat{A}=\hat{O}(\varepsilon^\nu)$, if
 \begin{align}
&\frac{\|\hat{A}u_s\|}{\|u_s\|}=O(\varepsilon^\nu),
\label{est-1}
\end{align}
where $\|u_s\|=\|u_s\|(t)=\sqrt{\int_{-\infty}^{\infty}u^2(x,t)dx}$ is  the  $L_2$-norm
of $u_s$.
\par
 From Eq. \eqref{est-1}, we  directly obtain the following estimates for the products and powers of the operators $\Delta x_s=x-X_s(t,\varepsilon)$ and $\hat{p}=\varepsilon\partial_x$:
   \begin{align}
&\frac{\| (\Delta x_s)^k\hat{p}^l u_s \|}{\|u_s \|}=O(\varepsilon^{(k+l)/2}),\quad
\quad k,l=0,1,2, \ldots ,
\label{est-1a}
\end{align}
and, in particular,
 \begin{align}
&  \Delta x_s =\hat{O}(\sqrt{\varepsilon}),\quad
 \hat{p}=\hat{O}(\sqrt{\varepsilon}),
 \label{est-1b}
\end{align}
just as in the case of the equation \eqref{fkpp} \cite{trsh2009,shtr2018,kulsh2024}.
\par
Note that we cannot directly obtain an estimate for the time derivative operator
 $D^{\alpha}_{F,t}$  in the equations \eqref{flsys1a}
within the class $\mathcal{P}_{(\alpha)t,s}^{\varepsilon}$ when the functional parameters $X_s(t,\varepsilon)$ and $R_s(t,\varepsilon)$ are arbitrary.
However, an estimate can be obtained for a ``prolonged'' time derivative operator
\begin{align}
&\hat{\mathcal{T}}^{\alpha}_{Fs}=\varepsilon D^{\alpha}_{F,t}+\big(D^{\alpha}_{F,t} X_s(t,\varepsilon) \big)\hat{p}- \big(D^{\alpha}_{F,t} R_s(t,\varepsilon)\big),
\label{toper-1}
\end{align}
which  accounts for both the structure of functions from the class $\mathcal{P}_{(\alpha)t,s}^{\varepsilon}$  and properties of the derivative $D^\alpha_{F,t}$  such as linearity, the Leibniz rule \eqref{leib-1}, and the properties of the chain rule \eqref{chrl-1}.
\par
By analogy with \eqref{est-1}, \eqref{est-1a}, \eqref{est-1b} for \eqref{toper-1}, we directly obtain the  estimate
\begin{align}
&\hat{\mathcal{T}}^{\alpha}_{Fs}=\hat{O}(\varepsilon).
\label{toper-1a}
\end{align}

\subsection{The moments}

For functions $u_s(x,t,\varepsilon)$ belonging to the class $\mathcal{P}_{(\alpha)t,s}^{\varepsilon}$, we  define  the moments
\begin{align}
&\mu_{u_s}(t,\varepsilon)=\int_{-\infty}^{\infty}u_s(x,t,\varepsilon)dx, \quad
x_{u_s}(t,\varepsilon)=\frac{1}{\mu_{u_s}(t,\varepsilon)}\int_{-\infty}^{\infty}x
u_s(x,t,\varepsilon)dx,\cr
&\alpha^{(l)}_{u_s}(t,\varepsilon)= \frac{1}{\mu_{u_s}(t,\varepsilon)}\int_{-\infty}^{\infty}\Delta x_s^l u_s(x,t,\varepsilon)dx,
\quad s=1,2,\ldots , K,\quad l=2,3,\ldots  ,
\label{mom-11}
\end{align}
which exist by virtue of the definitions \eqref{eq1} and \eqref{eq1a} of $\mathcal{P}_{(\alpha)t,s}^{\varepsilon}$.

Here, $\mu_{u_s}$ and $x_{u_s}$ are the zeroth-order and first-order moments, respectively, and
$\alpha^{(l)}_{u_s}$  is  the $l$-th central moment of $u_s(x,t,\varepsilon)$.
We also impose the following condition on the functional parameter $X_s(t,\varepsilon)$ of the  class $\mathcal{P}_{(\alpha)t,s}^{\varepsilon}$ \cite{trsh2009,shtr2018,kulsh2024}:
\begin{align}
&X_s(t,\varepsilon)=x_{u_s}(t,\varepsilon),
\label{mom1aa}
\end{align}
then  the function $u_s(x,t,\varepsilon)\in \mathcal{P}_{(\alpha)t,s}^{\varepsilon}$ will be  concentrated in the space neighborhood of the curve  $x= x_{u_s}(t,\varepsilon)$.
 The second moment $\alpha^{(2)}_{u_s}$ characterizes  the relative deviation of $u_s$  normalized to $\mu_{u_s}$.

To simplify the notation, we drop the dependence on the variables
 $t, \varepsilon$ and the  subscript $u_s$  in what follows, where this does not lead to confusion.
Specifically, we write:
$\mu_{u_s}(t,\varepsilon)=\mu_{u_s}(t)$, $x_{u_s}(t,\varepsilon)=x_{u_s}(t)$,
$\alpha^{(l)}_{u_s}(t,\varepsilon)=\alpha^{(l)}_{u_s}(t)$.
\par
From Eqs. \eqref{mom-11} and \eqref{mom1aa}, we derive the following estimates for the moments \cite{trsh2009,shtr2018}:
\begin{align}
\mu_{u_s}=O(1),
\quad \alpha^{(l)}_{u_s}=O(\varepsilon^{l/2}),\quad \varepsilon\to 0.
\label{mom1b}
\end{align}
and  setting $l=0,1,2$ in \eqref{mom-11}, we have
\begin{align}
\alpha^{(0)}_{u_s}=1,\quad \alpha^{(1)}_{u_s} =0 .
\label{mom1bb}
\end{align}

\subsection{Expansion of equation coefficients}

To construct asymptotic solutions to equation \eqref{flfkpp1}, we expand the coefficients
$a(x,t)$, $b(x,y,t)$  into formal power series in a neighborhood of the trajectory
$x=X_s(t,\varepsilon)$ \cite{shtr2018,kulsh2024}:
\begin{align}
&a(x,t)=\sum_{k=0}^{\infty}\frac{1}{k!}a_{(s)k}(t,\varepsilon)\Delta x_{s}^k,\quad
b(x,y,t)=\sum_{k,l=0}^{\infty}\frac{1}{k!l!}b_{(s,\bar{s})k,l}(t,\varepsilon)\Delta x_{s}^k\Delta y_{\bar{s}}^l,\label{expan-1aa}
\end{align}
where  $\Delta x_s=x-X_s(t,\varepsilon)$, $\Delta y_{\bar{s}}=y-X_{\bar{s}}(t,\varepsilon)$, and
\begin{align}
&a_{(s)k}(t,\varepsilon)=\frac{\partial^k a(x,t)}{\partial x^k}\Big|_{x={X_s}(t,\varepsilon)},\cr
& b_{(s,\bar{s})k,l}(t,\varepsilon)=\frac{\partial^{k+l} b(x,y,t)}{\partial x^k\partial y^l}\Big|_{
\begin{subarray}{c}
    x= X_{s}(t,\varepsilon)\\
    y= X_{\bar{s}}(t,\varepsilon)
 \end{subarray}} ,\quad b_{(s,s)k,l}(t,\varepsilon)=b_{(s)k,l}(t,\varepsilon).
\label{expan-1b}
\end{align}
For the coefficients of the expansions in \eqref{expan-1b}, we will use the simplified notations $a_{(s)k}(t,\varepsilon)=a_{(s)k}(t)$, $b_{(s)k,l}(t,\varepsilon)=b_{(s)k,l}(t)$, and
$b_{(s,\bar{s})k,l}(t,\varepsilon)=b_{(s,\bar{s})k,l}(t)$.
\par
The solution $u$ to equation \eqref{flfkpp1} within the semiclassical framework is characterized by its asymptotic expansion in powers of  $\sqrt{\varepsilon}$.
 The leading term and the first two corrections in this expansion describe the essential features of the solution with an accuracy of $O(\varepsilon^{3 /2})$, see \cite{trsh2009,shtr2018,shtr2019}. To construct explicit analytical expressions for these asymptotic terms, the zeroth-order moment $\mu_{u_s}$, the first-order moment  $x_{u_s}$, and the second-order moment $\alpha_{u_s}^{(2)}$ are computed, each with an accuracy of $O (\varepsilon^{3 /2})$.
\par
Given this, the analysis can be limited to equations for moments up to second order, as higher-order moments do not contribute to the solution at the desired level of accuracy. This simplification aligns with the semiclassical approach, where higher-order terms in the asymptotic series are typically negligible for practical purposes.
\par
We now derive the dynamical system for the moments defined in equations \eqref{mom-11} by employing equation \eqref{flfkpp1}, the expansions in \eqref{expan-1aa} and \eqref{expan-1b}, and the estimates provided in \eqref{mom1b}.

\section{The dynamical system of  moments}
\label{sect-momdyn}

Let us apply the derivative $D^\alpha_{F,t}$ to $\mu_{u_s}(t,\varepsilon)$ in Eq. \eqref{mom-11} and substitute $D^\alpha_{F,t}u_s(x,t, \varepsilon)$, where $u_s(x,t,\varepsilon)\in \mathcal{P}_{(\alpha)t,s}^{\varepsilon}$, into the right-hand side of the relation obtained from Eq. \eqref{flsys1a}. This yields
\begin{align}
&D^\alpha_{F,t}\mu_{u_s}(t,\varepsilon)=\chi_{F}(t)\int_{-\infty}^{\infty}A(x,t,u,\varepsilon)u_s(x,t,\varepsilon)dx,\label{mu-1a}\\
&A(x,t,u,\varepsilon)=a(x,t)-\varkappa\int\limits_{-\infty}^{\infty}b(x,y,t)\sum_{\bar{s}=1}^{K}
u_{\bar{s}}(y,t,\varepsilon)dy.
\label{mu-1aa}
\end{align}
Consider $A(x,t,u,\varepsilon)$ in a neighborhood of the trajectory $x=X_s(t,\varepsilon)$.
Expand $A(x,t,u,\varepsilon)$ in terms of $\Delta x_s$   and $\Delta y_{\bar{s}}$
using the expansions \eqref{expan-1aa} and \eqref{expan-1b}, along with the notation \eqref{mom-11}.
Then  for the case $A(x,t,u,\varepsilon)=A_s(x,t,u,\varepsilon)$,
we can express \eqref{mu-1aa} as a formal series
\begin{align}
&A_s(x,t,u,\varepsilon)=\sum_{k=0}^{\infty}\frac{1}{k!}(\Delta x_s)^k \Big[a_{(s)k}(t,\varepsilon)-
\varkappa\sum_{l=0}^{\infty}\frac{1}{l!}\sum_{\bar{s}=1}^{K}
b_{(s,\bar{s})k,l}(t,\varepsilon)
\mu_{u_{\bar{s}}}(t,\varepsilon) \alpha^{(l)}_{u_{\bar{s}}}(t,\varepsilon)\Big].
\label{mu-1ab}
\end{align}
In view of  estimates \eqref{est-1b}, \eqref{mom1b}, and  \eqref{mom1bb}, expansion \eqref{mu-1ab} can be written  accurate to $O(\varepsilon^{3/2})$ as
\begin{align}
&A_s(x,t,u,\varepsilon)=a_{(s)0}-
\varkappa\sum_{\bar{s}=1}^{K}\mu_{u_{\bar{s}}}\Big(b_{(s,\bar{s})0,0}+
\frac{1}{2}b_{(s,\bar{s})0,2}\alpha^{(2)}_{u_{\bar{s}}} \Big)
\cr
&+
\Delta x_s\Big( a_{(s)1}-\varkappa\sum_{\bar{s}=1}^{K}\mu_{u_{\bar{s}}}
b_{(s,\bar{s})1,0}\Big)
\cr
&+\frac{1}{2}(\Delta x_s)^2\Big(a_{(s)2} -\varkappa\sum_{\bar{s}=1}^{K}\mu_{u_{\bar{s}}}
b_{(s,\bar{s})2,0}\Big)+O(\varepsilon^{3/2}),
\label{mu-1ac}
\end{align}
where $a_{(s)k}=a_{(s)k}(t,\varepsilon)$, $b_{(s,\bar{s})k,l}=b_{(s,\bar{s})k,l}(t,\varepsilon)$; $k,l=0,1,2$; $\mu_{u_s}=\mu_{u_s}(t,\varepsilon)$, $\alpha_{u_s}^{(2)}=\alpha_{u_s}^{(2)}(t,\varepsilon)$.
Then from \eqref{mu-1ac} we have
\begin{align}
&\Delta x_s A_s(x,t,u,\varepsilon)=\Delta x_s\Big(a_{(s)0}-\varkappa\sum_{\bar{s}=1}^{K}\mu_{u_{\bar{s}}}b_{(s,\bar{s})0,0}\Big)
\cr
&+(\Delta x_s)^2\Big( a_{(s)1}-\varkappa\sum_{\bar{s}=1}^{K}\mu_{u_{\bar{s}}}b_{(s,\bar{s})1,0}\Big)+O(\varepsilon^{3/2}),\cr
&(\Delta x_s)^2 A_s(x,t,u,\varepsilon)=(\Delta x_s)^2\Big(a_{(s)0}-\varkappa\sum_{\bar{s}=1}^{K}\mu_{u_{\bar{s}}}b_{(s,\bar{s})0,0}\Big)+O(\varepsilon^{3/2}).
\label{mu-1ad}
\end{align}
Here, we take in mind that $\mu_{u_s}=O(1)$,  $\Delta x_s=O(\sqrt{\varepsilon})$,
$\alpha^{(2)}_{u_s} =O(\varepsilon)$.
\par
From \eqref{mu-1ac} and \eqref{mu-1ad} it follows directly
\begin{align}
&\int_{-\infty}^{\infty}A_s(u,x,t,\varepsilon)u_s(x,t,\varepsilon)dx=
\mu_{u_s}\Big(a_{(s)0}-\varkappa\sum_{\bar{s}=1}^{K}\mu_{u_{\bar{s}}}\big(b_{(s,\bar{s})0,0}+
\frac{1}{2}b_{(s,\bar{s})0,2}\alpha^{(2)}_{u_{\bar{s}}}\big)\Big)
\cr
&+ \frac{1}{2}\mu_{u_s}\alpha^{(2)}_{u_s}\Big(a_{(s)2}-
\varkappa\sum_{\bar{s}=1}^{K}\mu_{u_{\bar{s}}}b_{(s,\bar{s})2,0}\Big)+ O(\varepsilon^{3/2}),\label{eqj-a5}\\
&\int_{-\infty}^{\infty}\Delta x_s A_s(u,x,t,\varepsilon)u_s(x,t,\varepsilon)dx=
\mu_{u_s}\alpha_{u_s}^{(2)}\Big(a_{(s)1}-\varkappa\sum_{\bar{s}=1}^{K} \mu_{u_{\bar{s}}} b_{(s,\bar{s})1,0} \Big)+O(\varepsilon^{3/2}),\label{eqj-b5}\\
&\int_{-\infty}^{\infty}(\Delta x_s)^2 A_s(u,x,t,\varepsilon)u_s(x,t,\varepsilon)dx=
\mu_{u_s}\alpha_{u_s}^{(2)}\Big(a_{(s)0}-\varkappa \sum_{\bar{s}=1}^{K}\mu_{u_{\bar{s}}} b_{(s,\bar{s})0,0} \Big)+O(\varepsilon^{3/2}).
\label{eqj-c5}
\end{align}
From \eqref{mu-1a} and \eqref{mu-1ab} we obtain
\begin{align}
&D^\alpha_{F,t}\mu_{u_s}=\chi_{F}(t)\mu_{u_s}
\sum_{k=0}^{\infty}\frac{\alpha^{(k)}_{u_s}}{k!}\Big[
a_{(s)k}(t)-\varkappa \sum_{l=0}^{\infty}\frac{1}{l!}\sum_{\bar{s}=1}^{K}
b_{(s,\bar{s})k,l}(t)\mu_{u_{\bar{s}}}
\alpha^{(l)}_{u_{\bar{s}}}\Big].
\label{mu-2a}
\end{align}
Equation \eqref{mu-2a} and similar equations can be treated as approximations with a specified accuracy in $\varepsilon$, provided we restrict ourselves to a finite number of terms in the infinite sums of \eqref{mu-2a}, while considering the estimates \eqref{mom1b}.

By neglecting terms of order $O(\varepsilon^{3/2})$ and higher in equation \eqref{mu-2a}, and incorporating \eqref{eqj-a5}, we derive the following evolution equation for $\mu_{u_s}$, which includes moments up to the second-order:
\begin{align}
&D^{\alpha}_{F,t}\,\mu_{u_s}=\chi_{F}(t)\mu_{u_s}\Big[a_{(s)0}+\frac{1}{2}
\Big(a_{(s)2}- \varkappa \sum_{\bar{s}=1}^{K}b_{(s,\bar{s})2,0}\mu_{u_{\bar{s}}} \Big)\alpha^{(2)}_{u_s}
 \cr
&-\varkappa\Big(\sum_{\bar{s}=1}^{K}\mu_{u_{\bar{s}}}b_{(s,\bar{s})0,0}+
\frac{1}{2} \sum_{\bar{s}=1}^{K} b_{(s,\bar{s})0,2}\mu_{u_{\bar{s}}}\alpha^{(2)}_{u_{\bar{s}}}\Big)
\Big] + O(\varepsilon^{3/2}).
\label{mu-2b}
\end{align}
By applying the derivative operator $D_{F,t}^{\alpha}$  to the first-order moment $x_{u_s}=x_{u_s}(t,\varepsilon)$ and carrying out the necessary calculations, we arrive at the following equation:
\begin{align}
&D^{\alpha}_{F,t}\,x_{u_s}=\frac{1}{\mu_{u_s}}\chi_{F}(t)\int\limits_{-\infty}^{\infty}
\Delta x_{u_s} A(u,x,t,\varepsilon)u_s(x,t,\varepsilon)dx.
\label{momx-1a}
\end{align}
Then we have
\begin{align}
&D^{\alpha}_{F,t}\,x_{u_s}=\chi_{F}(t)\sum_{k=0}^{\infty}\frac{1}{k!}\alpha_{u_s}^{(k+1)} \Big(a_{(s)k}-\varkappa \sum_{l=0}^{\infty}\frac{1}{l!}\sum_{\bar{s}=1}^{K}b_{(s,\bar{s})k,l}\mu_{u_{\bar{s}}}
\alpha_{u_{\bar{s}}}^{(l)}\Big).
\label{momx-1bb}
\end{align}
Using expression \eqref{eqj-b5}, we obtain the following evolutionary equation for $x_{u_s}$, which includes moments up to the second-order:
\begin{align}
&D^{\alpha}_{F,t}\,x_{u_s}=\chi_{F}(t)\alpha_{u_s}^{(2)} \Big(a_{(s)1}-\varkappa\sum_{\bar{s}=1}^{K}\mu_{u_{\bar{s}}} b_{(s,\bar{s})1,0}\Big)+O(\varepsilon^{3/2}).
\label{momx-1b}
\end{align}
Similarly, the equations for higher-order moments are derived in the following form:
\begin{align}
&D^{\alpha}_{F,t}\,\alpha_{u_s}^{(n)}=\chi_{F}(t)\varepsilon n(n-1)\alpha_{u_s}^{(n-2)}+
\chi_{F}(t)\left\{\sum_{k=0}^{\infty}\frac{1}{k!}a_{(s)k}\left(
\alpha_{u_s}^{(k+n)}-\alpha_{u_s}^{(k)}\alpha_{u_s}^{(n)}-n\alpha_{u_s}^{(k+1)}\alpha_{u_s}^{(n-1)}\right)
\right.\cr
&\left.+\varkappa\sum_{k=0}^{\infty}\frac{1}{k!}\sum_{l=0}^{\infty}\frac{1}{l!}
\sum_{\bar{s}=1}^{K}b_{(s,\bar{s})k,l}\mu_{u_{\bar{s}}}\alpha_{u_{\bar{s}}}^{(l)}\left(
-\alpha_{u_s}^{(n+k)}+n\alpha_{u_s}^{(k+1)}\alpha_{u_s}^{(n-1)}+\alpha_{u_s}^{(k)}\alpha_{u_s}^{(n)}\right)
\right\}.
\label{malp-1bb}
\end{align}
By neglecting terms of the order $O(\varepsilon^{3/2})$ and higher in equation  \eqref{malp-1bb}, we derive the following evolution equation for the second-order moment $\alpha_{u_s}^{(2)}$:
\begin{align}
&D^{\alpha}_{F,t}\,\alpha_{u_s}^{(2)}=2\varepsilon \chi_{F}(t).
\label{malp-1b}
\end{align}
We observe that the dynamical system of moments, defined by equations \eqref{mu-2a}, \eqref{momx-1b}, and \eqref{malp-1b}, characterizes the localization trajectory of an asymptotic solution $u$ to equation \eqref{flfkpp1}.  This solution is found
with an accuracy of $O(\varepsilon^{3/2})$ within the class of functions $\mathcal{P}_{(\alpha)t,s}^{\varepsilon}$ of the form \eqref{eq1} (see, for example, \cite{trsh2009} and \cite{shtr2018}).
\par

\subsection{The fractal Einstein--Ehrenfest system of the second-order }

Consider equations \eqref{mu-2b}, \eqref{momx-1b}, and \eqref{malp-1b} as templates, and introduce a dynamical system where the variables are not moments of the solutions $u_s$ to the nonlinear system \eqref{flsys1a}.
\par
Let
\begin{align}
&\mathfrak{g}_s(t)=\big\{\mu_s(t),x_s(t),\alpha_s^{(2)}(t) \big\},
\label{fees1}
\end{align}
be a vector consisting of functions $\mu_s(t)$, $x_s(t)$, and $\alpha_s^{(2)}(t)$ which are not moments of any function $u_s(x,t,\varepsilon)$.  We  assume that the functions \eqref{fees1} are $F^{\alpha}$-differentiable in the sense of the definition \eqref{fdir-1}.
\par
Let us replace the moments $\mu_{u_s}(t,\varepsilon)$, $x_{u_s}(t,\varepsilon)$, $\alpha_{u_s}^{(2)}(t,\varepsilon)$ in equations \eqref{mu-2b}, \eqref{momx-1b}, \eqref{malp-1b}  with the corresponding functions \eqref{fees1}. Then  we obtain the following dynamical system for functions \eqref{fees1}:
\begin{align}
&D^{\alpha}_{F,t}\,\mu_{s}=\chi_{F}(t)\mu_{s}\Big[a_{(s)0}+\frac{1}{2}
\Big(a_{(s)2}- \varkappa \sum_{\bar{s}=1}^{K}b_{(s,\bar{s})2,0}\mu_{{\bar{s}}} \Big)\alpha^{(2)}_{s}
 \cr
&-\varkappa\Big(\sum_{\bar{s}=1}^{K}\mu_{\bar{s}}b_{(s,\bar{s})0,0}+
\frac{1}{2} \sum_{\bar{s}=1}^{K} b_{(s,\bar{s})0,2}\mu_{\bar{s}}\alpha^{(2)}_{\bar{s}}\Big)
\Big], \label{fees2a}\\
&D^{\alpha}_{F,t}\,x_{s}=\chi_{F}(t)\alpha_{s}^{(2)} \Big(a_{(s)1}-\varkappa\sum_{\bar{s}=1}^{K}\mu_{\bar{s}} b_{(s,\bar{s})1,0}\Big), \label{fees2b}\\
&D^{\alpha}_{F,t}\,\alpha_s^{(2)}=2\varepsilon \chi_{F}(t).
\label{fees2c}
\end{align}
We refer to equations \eqref{fees2a}, \eqref{fees2b}, and \eqref{fees2c} as the fractal Einstein--Ehrenfest system (FlEES) of the 2nd-order, following the terminology established in \cite{trsh2009,shtr2018,shtr2019}. We will also retain the estimates given in \eqref{mom1b} for the variables in \eqref{fees1}.
\par
Let the general solution of the system \eqref{fees2a}--\eqref{fees2c} be denoted by
\begin{align}
&\mathfrak{g}_s(t,\mathbf{C})=\big\{\mu_s(t,\mathbf{C}),x_s(t,\mathbf{C}),\alpha_s^{(2)}(t,\mathbf{C}) \big\},
\label{fees3}
\end{align}
where $\mathbf{C}$ represents a set of arbitrary integration constants, which corresponds to the entire system of  interacting quasiparticle equations.

  \subsection{On the ``prolongation'' of the fractal time derivative }

When constructing asymptotic solutions to the equations \eqref{flsys1a} within the classes \eqref{eq1a}, it is more convenient to apply estimates of the operators \eqref{est-1} defined not {\it on arbitrary} functions $u_s$ from the class \eqref{eq1a}, but {\it on solutions} $u_s$ of the equations \eqref{flsys1a} with $u_s\in \mathcal{P}_{(\alpha)t,s}^{\varepsilon}$.
\par
In this case, the estimates \eqref{est-1b} for the operators $\Delta x_s$ and $\hat{p}$ will remain unchanged.
However, the functional parameter $X_s(t,\varepsilon)$, as determined by \eqref{mom1aa}, will be governed by the equation \eqref{momx-1b} (or \eqref{momx-1bb}), from which it follows that $D^{\alpha}_{F,t}X_s(t,\varepsilon)=O(\varepsilon)$. Thus, the operator $D^{\alpha}_{F,t}X_s(t,\varepsilon)\hat{p}$, which is included in the ``prolonged'' operator
$\hat{\mathcal T}^{\alpha} _{Fs}$ of the form \eqref{toper-1}, receives the estimate
$\hat{O}(\varepsilon^{3/2})$, which exceeds the estimate \eqref{toper-1a} for the operator
$\hat{\mathcal T }^{\alpha}_{Fs}$.
This allows us to simplify the ``prolonged'' operator $\hat{\mathcal T}^{\alpha}_{Fs}$
by removing the operator $D^{\alpha}_{F,t}X_s(t,\varepsilon)\hat{p}$, without violating the estimate $\hat{O}(\varepsilon)$ for $\hat{\mathcal T}^{\alpha}_{Fs}$.
\par
Additional simplifications are obtained by imposing the following condition on the functional parameter $R_s(t,\varepsilon)$ from the class \eqref{eq1a}:
\begin{align}
&D^{\alpha}_{F,t}R_s=\varepsilon\frac{D^{\alpha}_{F,t}\mu_{u_s}}{\mu_{u_s}}.
\label{r-1}
\end{align}
Thus, in what follows, instead of the operator $\hat{\mathcal T}^{\alpha}_{F,t}$ in the form of equation \eqref{toper-1}, we will use the simplified ``prolonged'' fractal time derivative operator
 \begin{align}
&\hat{T}_{Fs}^{\alpha}=\varepsilon D^{\alpha}_{F,t}-\varepsilon\frac{D^{\alpha}_{F,t}\mu_{u_s}}{\mu_{u_s}}
\label{stopr-1a}
\end{align}
with the estimate
 \begin{align}
&\hat{T}_{Fs}^{\alpha}=\hat{O}(\varepsilon).
\label{stopr-1b}
\end{align}

\section{Associated fractal linear equations}
\label{sect-ale}

Consider the nonlinear system of equations \eqref{flsys1a} along with the expansions \eqref{expan-1aa},  \eqref{expan-1b}, \eqref{mu-1ab} 
in the following form:
\begin{align}
&-D^\alpha_{F,t}u_s(x,t,\varepsilon)+\chi_F(t)\Big\{\varepsilon u_{s\,xx}(x,t,\varepsilon)
\cr
&+\sum_{k=0}^{\infty}\frac{1}{k!}(\Delta x_s)^k\Big(a_{(s)k}(t,\varepsilon) -\varkappa\sum_{l=0}^{\infty}
\frac{1}{l!}\sum_{\bar{s}=1}^{K}b_{(s,\bar{s})k,l}(t,\varepsilon) \mu_{\bar{s}}\alpha_{u_{\bar{s}}}^{(l)}\Big)u_s(x,t,\varepsilon)\Big\}=0.
\label{treq-1a}
\end{align}
Let us derive the linear system of equations associated with the approximate nonlinear system \eqref{treq-1a}, where a finite number of terms are retained in the formal sums, based on the estimates \eqref{est-1a}, \eqref{toper-1a}, and \eqref{mom1b}.
\par
To achieve this, we restrict ourselves to moments up to second-order and replace the moments $\mu_{u_s}$, $x_{u_s}$, and $\alpha_{u_s}^{(2)}$ in equation \eqref{treq-1a} with the corresponding functions $\mu_s(t,\mathbf{C})$, $x_s(t,\mathbf{C})$, and $\alpha^{(2)}_s(t,\mathbf{C})$  from the general solution \eqref{fees3} of the fractal Einstein--Ehrenfest system.
This yields  the following system of linear equations, parameterized by the arbitrary integration constants $\mathbf{C}$:
\begin{align}
&\hat{L}_s(x,t,\varepsilon,\mathbf{C})v_s(x,t,\varepsilon,\mathbf{C})=0,
\label{fale-1}
\end{align}
where
\begin{align}
& \varepsilon \hat{L}_s(x,t,\varepsilon,\mathbf{C})\equiv
-\hat{T}^{\alpha}_{Fs}-\varepsilon\frac{D^{\alpha}_{F,t}\mu_s(t,\mathbf{C})}{\mu_s(t,\mathbf{C})}
\cr
&+\chi_F(t)\left(\hat{p}^2+\varepsilon\sum_{k=0}^{2}
\frac{1}{k!}(\Delta x_s)^k\Big( a_{(s)k}-\varkappa\sum_{l=0}^{2}\frac{1}{l!}\sum_{\bar{s}=1}^{K}b_{(s,\bar{s})k,l}
\mu_{\bar{s}}(t,\mathbf{C})\alpha_{\bar{s}}^{(l)}(t,\mathbf{C}) \Big)\right).
\label{fale-1b}
\end{align}
Here, $v(x,t,\varepsilon,\mathbf{C})$ is a solution to the equation \eqref{fale-1b},
and $a_{(s)k}=a_k(t,\varepsilon,\mathbf{C})$, $b_{(s,\bar{s})k,l}=b_{k,l}(t,\varepsilon,\mathbf{C})$ are
given by \eqref{expan-1b} where $\mu_s=\mu_s(t,\varepsilon,\mathbf{C})$ and $x_s(t,\mathbf{C})=X_s(t)=X_s(t,\varepsilon,\mathbf{C})$ are  the components of the general solution to the FlEES.
The operator $\hat{T}^\alpha_{Fs}=$ $\varepsilon D^{\alpha}_{F,t}-\varepsilon D^{\alpha}_{T,t}\mu_s/\mu_s$ is given by \eqref{stopr-1a}, where $\mu_{u_s}$ is replaced by $\mu_s(t,\varepsilon,\mathbf{C})$.
\par
We refer to the linear equation \eqref{fale-1b} as {\it the associated fractal linear equation} (AFlLE) for the fractal Fisher--KPP equation \eqref{flsys1a} associated with the $s$-quasiparticle.
\par
Using \eqref{mu-2a}, we rewrite equation \eqref{fale-1b} as
\begin{align}
& \varepsilon \hat{L}_s(x,t,\varepsilon,\mathbf{C})=
-\hat{T}^{\alpha}_{Fs}
+\chi_F(t)\Big\{\hat{p}^2+\varepsilon\Delta x_s\big[a_{(s)1}-\varkappa\sum_{\bar{s}=1}^{K}b_{(s,\bar{s})1,0}\mu_{\bar{s}}\big]
\cr
&+\varepsilon\frac{1}{2}\big[(\Delta x_s)^2-\alpha_{s}^{(2)}\big]
\Big( a_{(s)2}-\varkappa \sum_{\bar{s}=1}^{K}b_{(s,\bar{s})2,0}
\mu_{\bar{s}}(t,\mathbf{C}) \Big)\Big\}.
\label{fale-1d}
\end{align}

\subsubsection{The linear operator $\varepsilon\hat{L}_s$ expansion  accurate to $O(\varepsilon^{5/2})$}

Let us expand the operator $\varepsilon \hat{L}_s(x,t,\varepsilon,\mathbf{C})$
of the linear associated equation \eqref{fale-1}, \eqref{fale-1b}
given by \eqref{fale-1d} taking into account
estimates \eqref{est-1a}, \eqref{est-1b}, \eqref{mom1b}, and \eqref{stopr-1a}, \eqref{stopr-1b}. Then we have
\begin{align}
&\varepsilon \hat{L}_s(x,t,\varepsilon,\mathbf{C})=
 \hat{L}_{s(0)}(t,\varepsilon,\mathbf{C})+ \hat{L}_{s(1)}(t,\varepsilon,\mathbf{C})+ \hat{L}_{s(2)}(t,\varepsilon,\mathbf{C})+\hat{O}(\varepsilon^{5/2}),\cr
\label{Lprr-1a}
\end{align}
 where $\hat{L}_{s(0)}(t,\varepsilon,\mathbf{C})=\hat{O}(\varepsilon)$,
 $\hat{L}_{s(1)}(t,\varepsilon,\mathbf{C})=\hat{O}(\varepsilon^{3/2})$,
 $\hat{L}_{s(2)}(t,\varepsilon,\mathbf{C})=\hat{O}(\varepsilon^2)$, and
\begin{align}
&\hat{L}_{s(0)}(t,\varepsilon,\mathbf{C})= -\hat{T}^{\alpha}_{Fs}+\chi_F(t)\hat{p}^2 ,\label{Lprr-2a}\\
&\hat{L}_{s(1)}(t,\varepsilon,\mathbf{C})=\chi_F(t) \big(a_{(s)1}-\varkappa\sum_{\bar{s}}^{K} b_{(s,\bar{s})1,0}\mu_{\bar{s}}\big)\varepsilon \Delta x_s  ,\label{Lprr-2b}\\
& \hat{L}_{s(2)}(t,\varepsilon,\mathbf{C})=\chi_F(t)\frac{1}{2}
\big(a_{(s)2}-\varkappa\sum_{\bar{s}}^{K} b_{(s,\bar{s})2,0}\mu_{\bar{s}})\big)
\varepsilon\big((\Delta x_s)^2-\alpha^{(2)}_s\big).
\label{Lprr-2c}
\end{align}
We seek trajectory-concentrated solutions $v_s(t,x,\varepsilon,\mathbf{C})\in$
$\mathcal{P}_{(\alpha)t,s}^{\varepsilon}$ of the TCF \eqref{eq1a} for the linear associated equation (\ref{fale-1}) in the form
\begin{equation}
\label{eq19}
v_s(x,t,\varepsilon,\mathbf{C})=v_s^{(0)}(x,t,\varepsilon,\mathbf{C})+\sqrt{\varepsilon}
v_s^{(1)}(x,t,\varepsilon,\mathbf{C})+\varepsilon v_s^{(2)}(x,t,\varepsilon,\mathbf{C})+
\ldots ,
\end{equation}
where $v^{(k)}_{s}(x,t,\varepsilon,\mathbf{C})$ has the estimate $O(\varepsilon^0)$ over the time interval $t\in [0,T]$.

Substituting expansion \eqref{eq19} into  \eqref{fale-1} and \eqref{Lprr-1a}, we obtain the following system of recurrent equations:
\begin{align}
& \hat{L}_{s(0)}(t,\varepsilon,\mathbf{C})v_s^{(0)}=0,\label{eq23a}\\
& \sqrt{\varepsilon}\hat{L}_{s(0)}(t,\varepsilon,\mathbf{C})v_s^{(1)}+
\hat{L}_{s(1)}(t,\varepsilon,\mathbf{C})v_s^{(0)}=0, \label{eq23b}\\
& \varepsilon\hat{L}_{s(0)}(t,\varepsilon,\mathbf{C})v_s^{(2)}+
\sqrt{\varepsilon}\hat{L}_{s(1)} (t,\varepsilon,\mathbf{C})v_s^{(1)}+
\hat {L}_{s(2)}(t,\varepsilon,\mathbf{C})v_s^{(0)}=0 ,\label{eq23c}
\end{align}
where the operators $\hat{L}_{s(k)}$, $k=0,1,2$, are defined in equations \eqref{Lprr-2a}--\eqref{Lprr-2c}.

\section{The Cauchy problem}
\label{sect-cauch}

The Cauchy conditions for equations \eqref{flsys1a} in the function classes defined by \eqref{eq1a} are given by
\begin{align}
&u_{s}(x,t,\varepsilon)|_{t=0}=\varphi_{s}(x,\varepsilon),\quad s=1,\ldots , K,\cr
&u_{s}(x,t,\varepsilon)\in \mathcal{P}_{(\alpha)t,s}^{\varepsilon}, \quad
\varphi_{s}(x,\varepsilon)\in \mathcal{P}_{(\alpha)0,s}^{\varepsilon}=\mathcal{P}_{(\alpha)t,s}^{\varepsilon}|_{t=0}.
\label{cauch-1a1}
\end{align}
Then the corresponding Cauchy conditions for FlEES \eqref{fees2a}, \eqref{fees2b}, and \eqref{fees2c} are defined as
\begin{align}
&\mu_{s}(t,\varepsilon)|_{t=0}=\mu_{s}(\varepsilon,\varphi_s),\quad x_{s}(t,\varepsilon)|_{t=0}=x_{s}(\varepsilon,\varphi_s),\quad  \alpha_{s}^{(2)}(t,\varepsilon)|_{t=0}= \alpha_{s}^{(2)}(\varepsilon,\varphi_s),
\label{cauch-1b1}
\end{align}
where  $\mu_{s}(\varepsilon,\varphi_s)$, $x_{s}(\varepsilon,\varphi_s)$, and $\alpha_{s}^{(2)}(\varepsilon,\varphi_s)$ are given by \eqref{mom-11} for
$t=0$ and $u_{s}=\varphi_{s}$.
\par
Let us impose the conditions
\begin{align}
&\mu_{s}(t,\mathbf{C},\varepsilon)|_{t=0}=\mu_{s}(\varepsilon,\varphi_s),\quad x_{s}(t,\mathbf{C},\varepsilon)|_{t=0}=x_{s}(\varepsilon,\varphi_s),\cr  &\alpha_{s}^{(2)}(t,\mathbf{C},\varepsilon)|_{t=0}= \alpha_{s}^{(2)}(\varepsilon,\varphi_s)
\label{cauch-1cc}
\end{align}
on  the integration constants $\mathbf{C}$  in the general solution \eqref{fees3}
and take into account \eqref{cauch-1b1}.
Under these conditions, the integration constants $\mathbf{C}$ become functionals $\mathbf{C}[\vec{\varphi}]$,  which depend on the initial functions $\vec{\varphi}=(\varphi_1, \varphi_2, \ldots , \varphi_K)$ of all quasiparticles, where  $\varphi_s$ (for $s=1,\ldots ,K$) represents the initial function of the $s$-th  quasiparticle.
\par
Following the works of \cite{trsh2009, shtr2018, kulsh2024}, and considering the FlEES equations \eqref{fees2a}--\eqref{fees2c}, the expansions of the operators \eqref{Lprr-1a}--\eqref{Lprr-2c}, and the solution given by \eqref{eq19}, it can be directly found  that the approximate asymptotic solution to the system of equations \eqref{flsys1a}, subject to the initial conditions \eqref{cauch-1a1}, is given as
\begin{align}
&u_{s}(x,t,\varepsilon)=v_{s}(x,t,\mathbf{C}[\vec{\varphi}],\varepsilon)+O(\varepsilon^{3/2}),\quad s=1, \ldots ,K.
\label{cauch-2a}
\end{align}
Here,
\begin{align}
&v_{s}(x,t,\mathbf{C},\varepsilon)=\sum_{k=0}^{2} v_{s}^{(k)}(x,t,\mathbf{C},\varepsilon),
\label{cauch-2b}
\end{align}
 and the functions $v_{s}^{(k)}(x,t,\mathbf{C},\varepsilon)$ are determined by the equations
\eqref{eq23a}--\eqref{eq23c},  where $\mathbf{C}$ represents arbitrary constants, and the solution depends on the specified initial conditions
\begin{align}
& v_{s}^{(0)}(x,t,\mathbf{C},\varepsilon)|_{t=0}=\varphi_{s}(x,\varepsilon),\quad
v_{s}^{(1)}(x,t,\mathbf{C},\varepsilon)|_{t=0}=
v_{s}^{(2)}(x,t,\mathbf{C},\varepsilon)|_{t=0}=0.
\label{cauch-2c}
\end{align}
The accuracy of $O(\varepsilon^{3/2})$ in equation \eqref{cauch-2a} arises from the fact that we have restricted ourselves to the second-order FlEES \eqref{fees2a}--\eqref{fees2c} and the corresponding expansions in equations \eqref{Lprr-1a} and \eqref{eq19}.

\subsubsection{The Green function}

Let us construct asymptotic solutions to the equations \eqref{eq23a}, \eqref{eq23b}, and \eqref{eq23c} using the Green function and the Duhamel integral.
\par
We define the Green function $G_{s(0)}(x,y,t,t_0,\varepsilon,\mathbf{C})$ for the equation \eqref{eq23a} by the conditions
\begin{align}
&\hat{L}_{s(0)}(t,\varepsilon,\mathbf{C})G_{s(0)}(x,y,t,t_0,\varepsilon,\mathbf{C})=0,
\label{grn}\\
&G_{s(0)}(x,y,t_0,t_0,\varepsilon,\mathbf{C})=\delta(x-y).
\label{grn1}
\end{align}
For brevity we denote $G_{s(0)}(x,y,t,t_0,\varepsilon,\mathbf{C})=G_{s(0)}(x,y,t,t_0)$.
\par
Consider the Cauchy conditions
\begin{align}
&v_s^{(0)}(x,t,\varepsilon,\mathbf{C})|_{t=t_0}=
v_s^{(0)}(x,t_0,\varepsilon,\mathbf{C})=\varphi_s(x,\varepsilon,\mathbf{C}),\label{cgrn-1a}\\
&v_s^{(1)}(x,t,\varepsilon,\mathbf{C})|_{t=t_0}=v_s^{(2)}(x,t,\varepsilon,\mathbf{C})|_{t=t_0}=0,
\label{cgrn-1b}
\end{align}
where the initial function $\varphi_s(x,\varepsilon,\mathbf{C})$ belongs to the class defined in \eqref{eq1} at $t=t_0$. For generality, $\varphi_s$ may also depend on arbitrary constants  $\mathbf{C}$.
\par
The solution to the Cauchy problem for the equations \eqref{eq23a}, \eqref{eq23b}, and \eqref{eq23c}, subject to the conditions \eqref{cgrn-1a} and \eqref{cgrn-1b}, is given by the following expressions:
\begin{align}
&v_s^{(0)}(x,t,\varepsilon,\mathbf{C})=
\int_{-\infty}^{\infty}G_{s(0)}(x,y,t,t_0,\varepsilon,\mathbf{C})
v_s^{(0)}(y,t_0,\varepsilon,\mathbf{C})dy,\label{cauch-1a}\\
&v_s^{(1)}(x,t,\varepsilon,\mathbf{C})=\varepsilon^{-3/2}\int_{t_0}^{t}d^{\alpha}_{F}\tau
\int_{-\infty}^{\infty}G_{s(0)}(x,y,t,\tau,\varepsilon,\mathbf{C})
\hat{L}_{s(1)}(\tau,\mathbf{C})v_s^{(0)}(y,\tau,\varepsilon,\mathbf{C})dy,
\label{cauch-1b}\\
&v_s^{(2)}(x,t,\varepsilon,\mathbf{C})=
\int_{t_0}^{t}d^{\alpha}_{F}\tau
\int_{-\infty}^{\infty}G_{s(0)}(x,y,t,\tau,\varepsilon,\mathbf{C})\left[
\varepsilon^{-3/2}\hat{L}_{s(1)}(\tau,\mathbf{C})v_s^{(1)}(y,\tau,\varepsilon,\mathbf{C})
\right.\cr
&\left.+\varepsilon^{-5/2}\hat{L}_{s(2)}(\tau,\mathbf{C})v_s^{(0)}(y,\tau,\varepsilon,\mathbf{C})dy
\right].
\label{cauch-1c}
\end{align}
To simplify the notation, we will omit the explicit dependence on $\mathbf{C}$ in functions where doing so does not lead to ambiguity.
\par
In the Duhamel integrals \eqref{cauch-1b} and \eqref{cauch-1c}, we used the relationship between the fractal differentiation \eqref{fdir-1} and integration \eqref{int-2a}, \eqref{int-2aa}, see \cite{prgn2009}.
\par
The sum of the first three terms in the expansion \eqref{eq19}, with the functions $v^{(0)}$, $v^{(1)}$, and $v^{(2)}$ given by \eqref{cauch-1a}--\eqref{cauch-1c}, yields the asymptotic solution accurate to $O(\varepsilon^{3/2})$  for the linear associated equation \eqref{fale-1}.
\par
Let us substitute the explicit form of the operator $\hat{L}_{s(0)}$ from \eqref{Lprr-2a} and $\hat{T}^\alpha_{Fs}$ from \eqref{stopr-1a}  into equation \eqref{grn}. Then the Green function $G_{s(0)}(x,y,t,t_0)$, defined by the conditions \eqref{grn} and \eqref{grn1}  reads
\begin{align}
&\left(-\varepsilon D^{\alpha}_{F,t}+\varepsilon\frac{D^{\alpha}_{F,t}\mu_s}{\mu_s}+ \chi_F(t)(\varepsilon \partial_{x})^2
\right)G_{s(0)}(x,y,t,t_0)=0,
\label{grn-2}\\
&G_{s(0)}(x,y,t_0,t_0)=\delta(x-y).
\label{grn-2a}
\end{align}
The solution to the equations \eqref{grn-2} and \eqref{grn-2a} is obtained in the same way as for the case involving  the usual first-order time derivative, and it has the form (see also \cite{prgn2009}):
\begin{align}
&G_{s(0)}(x,y,t,t_0)=\frac{1}{2\sqrt{\pi \varepsilon\big(S_F(t)-S_F(t_0)\big)}}
\exp\left(-\frac{(x-y)^2}{4\varepsilon\big(S_F(t)-S_F(t_0)\big)}+\frac{R_s(t)-R_s(t_0)}{\varepsilon}\right).
\label{grn-2b}
\end{align}
For $D^{\alpha}_{F,t}R_s(t)=\varepsilon\frac{D^{\alpha}_{F,t}\mu_s(t)}{\mu_s(t)}$, we can write \begin{align*}
&G_{s(0)}(x,y,t,t_0)=\frac{1}{2\sqrt{\pi \varepsilon\big(S_F(t)-S_F(t_0)\big)}}
\frac{\mu_s(t)}{\mu_s(t_0)}\exp\left(-\frac{(x-y)^2}{4\varepsilon\big(S_F(t)-S_F(t_0)\big)}\right).
\end{align*}

\section{Example}
\label{exam-sol1}

Consider an example of constructing an asymptotic solution accurate to $O(\varepsilon^{3/2})$ in the form of quasiparticles (see Eq. \eqref{sol1a}) for the Fisher--KPP equation (Eq. \eqref{flfkpp1}) with a fractal time derivative $D^\alpha_{F,t}$ on the interval $t \in [0,1]$, where $F \subset [0,1] $, $F=C^\alpha$ is the Cantor set of the Hausdorff dimension $\alpha$. This example illustrates the general approach outlined in Sections \ref{sect1}--\ref{sect-cauch}. For simplicity, we restrict our analysis to the case of two interacting quasiparticles, $K=2$ in \eqref{sol1a}, where $s=1,2$. The quasiparticle functions $u_s(x,t,\varepsilon)$ are governed by the QDS  (Eqs. \eqref{flsys1a}), and the initial conditions are chosen in the form of Gaussian wave packets
\begin{align}
& \varphi_{s}(x,\varepsilon)=(\varepsilon)^{-1/2}N_s\exp\left( -\frac{\big(x-x_s(0)\big)^2}{2\varepsilon \sigma^2_{s}}\right),
\label{exam1}
\end{align}
which are localized in the  neighborhood of $x=x_s(0)$ for the $s$-th quasiparticle, with a normalization parameter $N_s$.  The parameter $\sigma_{s} (>0)$ characterizes the width of the Gaussian distribution.
We also choose the coefficients $b(x,y,t)$ in Gaussian form and $a(x,t)$ as a constant in equation \eqref{flfkpp1}:
\begin{align}
b(x,y,t)=b(x-y)=b_0\exp\Big[-\frac{(x-y)^2}{\xi^2}\Big],\quad a(x,t)=const=a.
\label{examp-B}
\end{align}
Here, $b_0$ represents the amplitude of the influence function $b(x,y,t)$, and $\xi(>0)$
 characterizes the width of the distribution given in equation \eqref{examp-B}. The present example can be compared to a similar case in \cite{kulsh2024} for the parameter $\alpha =1$ in the QDS described by equation \eqref{flsys1a}, which corresponds to a first-order time derivative.
\par
The asymptotic solution \eqref{sol1a} of the equation \eqref{flfkpp1}, accurate to $O(\varepsilon^{3/2})$, for two quasiparticles $u_s(x,t,\varepsilon)$ ($s=1,2$), as described by \eqref{cauch-2a}, \eqref{cauch-2b}, and \eqref{cauch-2c}, is given by
\begin{equation}
	\label{solEx}
 u(t,x,\varepsilon)=u_1(t,x,\varepsilon)+u_2(t,x,\varepsilon)=
  v(t,x,\mathbf{C}[\vec{\varphi}],\varepsilon)+O(\varepsilon^{3/2}),
\end{equation}
where
\begin{eqnarray}
\label{reshexamp}
&v(x,t,\mathbf{C}[\vec{\varphi}],\varepsilon)=
v_1^{(0)}(x,t,\mathbf{C}[\vec{\varphi}],\varepsilon)+
v_2^{(0)}(x,t,\mathbf{C}[\vec{\varphi}],\varepsilon)+
\varepsilon^{1/2}\big(v_1^{(1)}(x,t,\mathbf{C}[\vec{\varphi}],\varepsilon)\cr
&+v_2^{(1)}(x,t,\mathbf{C}[\vec{\varphi}],\varepsilon)\big)+\varepsilon\big(v_1^{(2)}(x,t,\mathbf{C}[\vec{\varphi}],\varepsilon)+v_2^{(2)}(x,t,\mathbf{C}[\vec{\varphi}],\varepsilon)\big).
\end{eqnarray}
\par
The functions $v_s^{(k)}(x,t,\varepsilon, \mathbf{C})$, $s=1,2$, $k=0,1,2$,  in \eqref{reshexamp} are given in the form \eqref{cauch-1a}--\eqref{cauch-1c}. The integration constants $\mathbf{C}$, which appear in the general solution \eqref{fees3} of the FlEES equations \eqref{fees2a}--\eqref{fees2c}, are determined by the conditions \eqref{cauch-1cc}.
\par
To write explicit expressions for the functions $v_s(x,t,\varepsilon, \mathbf{C})$, we introduce the notation
\begin{align}
&\Sigma_s(t,\alpha)=2 S^{\alpha}_{F}(t)+\sigma_s^2,\quad\cr
&\Delta R_s=R_s(t,\varepsilon)-R_s(0,\varepsilon).
\label{ex-fn-1}
\end{align}
Note that $\Sigma_s(t,\alpha)\big|_{t=0}=\sigma_s^2$, and the function $R_s(t,\varepsilon)$ is determined from the equation \eqref{r-1}.
\par
Using the formulas \eqref{cauch-1a}--\eqref{cauch-1c} and \eqref{grn-2b}, we obtain
\vspace*{-5mm}
\begin{align}
&v_s^{(0)}(x,t,\varepsilon)=
\varepsilon^{-1/2}N_s \exp\left( \frac{\Delta R_s}{\varepsilon}\right)
\exp\left(-\frac{\big( x-x_s(0))\big)^2}{2\varepsilon\Sigma_s(t,\alpha)}\right), \label{exam2a}
\end{align}
\vspace*{-1cm}
\begin{align}
&v_s^{(1)}(x,t,\varepsilon)=\frac{\sigma_s}{\varepsilon^{1/2}
\sqrt{\Sigma_s(t,\alpha)^3}} v^{(0)}_s(x,t,\varepsilon)
\cr
&\times \int^{t}_{0} d^{\alpha}_{F}\tau k_{(s)1}(\tau)\left[ \Sigma_s(\tau,\alpha)\big(x-x_s(0)\big)
-\Sigma_s(t,\alpha)\big(x_s(\tau)-x_s(0)\big)
 \right],  \label{exam2b}
\end{align}
\vspace*{-5mm}
\begin{align}
&v_s^{(2)}(x,t,\varepsilon)=\frac{\sigma_s}{\varepsilon \Sigma_s(t,\alpha)^{5/2}} v_s^{(0)}(x,t,\varepsilon)
\cr
&\times \Big\{\int_{0}^{t}d^{\alpha}_{F}\tau_2 k_{(s)1}(\tau_2)\int_{0}^{\tau_2}
d^{\alpha}_{F} \tau_1 k_{(s)1}(\tau_1)\Big[\Sigma_s(\tau_2,\alpha)\Sigma_s(\tau_1,\alpha)
\big(x-x_s(0)\big)
\cr
&-\Sigma_s(t,\alpha)\Big(\Sigma_s(\tau_2,\alpha)\big(x_s(\tau_1)-x_s(0)\big)+\Sigma_s(\tau_1,\alpha)
\big(x_s(\tau_2)-x_s(0)\big)\Big)\big(x-x_s(0)\big)
\cr
&+\Sigma_s(t,\alpha)\Big(2\varepsilon S^{\alpha}_{F}(t) \Sigma_s(\tau_1,\alpha)-4\varepsilon S^{\alpha}_{F}(\tau_1) S^{\alpha}_{F}(\tau_2)+ \Sigma_s(t,\alpha)\big(x_s(\tau_2)-x_s(0)\big)\big(x_s(\tau_1)-x_s(0)\big)\Big) \Big] 
\cr
&+\int_{0}^{t}d^{\alpha}_{F}\tau_2 k_{(s)2}(\tau_2)\Big[\Sigma_s(\tau_2,\alpha)\big(x-x_s(0)\big)^2-2\Sigma_s(t,\alpha)\Sigma_s(\tau_2,\alpha)
\big(x_s(\tau_2)-x_s(0)\big)\big(x-x_s(0)\big)
\cr
&+\Sigma_s(t,\alpha)\Big(-\alpha^{(2)}_s(\tau_2)\Sigma_s(t,\alpha)+2\varepsilon S_{F}^{\alpha}(t)\Sigma_s(\tau_2,\alpha)-2\varepsilon S_{F}^{\alpha}(\tau_2)\Sigma_s(\tau_2,\alpha)\cr
&+\Sigma_{F}^{\alpha}(t)\big(x_s(\tau_2)-x_s(0)\big)^2\Big)\Big]\Big\}.
\label{exam2c}
\end{align}
\vspace*{-5mm}
Here,  $k_{(s)1}(t)=a_{(s)1}(t)-\varkappa h_{(s)1}(t)$, $k_{(s)2}(t)=\frac{1}{2} a_{(s)2}(t)-\frac{1}{2}\varkappa h_{(s)2}(t)$, $h_{(s)k}=h^{(2)}_{(s)k}$,  and
\begin{equation*}
h^{(M)}_{(s)k}=\sum_{l-0}^{M-k}\frac{1}{l!}\sum_{\bar{s}=1}^{2}b_{(s,\bar{s})k,l}\mu_{\bar{s}}^{(M)}\alpha_{\bar{s}}^{(M)(l)}.
\end{equation*}
The FlEES, given by equations \eqref{fees2a}--\eqref{fees3}, takes the following form in this case:
\begin{align}
&D_{F,t}^{\alpha}\mu_1(t)=\Big(a\mu_1(t)-\varkappa\mu_1(t)\Big(\mu_1(t)+
\mu_2(t)b\big(x_1(t)-x_2(t)\big)-2\frac{\mu_1(t)\alpha_1^{(2)}(t)}{\xi^2}
\cr
&+\mu_2(t)\Big[2\frac{\big(x_1(t)-x_2(t)\big)^2}{\xi^4}-\frac{1}{\xi^2}\Big] b\big(x_1(t)-x_2(t)\big)\big(\alpha_1^{(2)}(t)+\alpha_2^{(2)}(t)\big)
\cr
&-3\frac{\mu_1(t)\big(\alpha_1^{(2)}(t)\big)}{\xi^4}+\Big[\frac{3}{\xi^4}
-12\frac{\big(x_1(t)-x_2(t)\big)^2}{\xi^6}+4\frac{\big(x_1(t)-x_2(t)\big)^4}{\xi^8}\Big]
\cr
&\times b\big(x_1(t)-x_2(t)\big)\mu_2(t)\alpha_1^{(2)}(t)\alpha_2^{(2)}(t)\Big),\cr
&D_{F,t}^{\alpha}\mu_2(t)=\Big(a\mu_2(t)-\varkappa\mu_2(t)\Big(\mu_2(t)+
\mu_1(t)b\big(x_1(t)-x_2(t)\big)-2\frac{\mu_2(t)\alpha_2^{(2)}(t)}{\xi^2}
\cr
&+\mu_1(t)\Big[2\frac{\big(x_1(t)-x_2(t)\big)^2}{\xi^4}-\frac{1}{\xi^2}\Big] b\big(x_1(t)-x_2(t)\big)\big(\alpha_1^{(2)}(t)+\alpha_2^{(2)}(t)\big)
\cr
&-3\frac{\mu_2(t)\big(\alpha_2^{(2)}(t)\big)}{\xi^4}+\Big[\frac{3}{\xi^4}-12\frac{\big(x_1(t)-x_2(t)\big)^2}{\xi^6}+4\frac{\big(x_1(t)-x_2(t)\big)^4}{\xi^8}\Big]
\cr
&\times b\big(x_1(t)-x_2(t)\big)\mu_1(t)\alpha_1^{(2)}(t)\alpha_2^{(2)}(t)\Big),
\label{eesexam1b}\\
&D_{F,t}^{\alpha}x_1(t)=\chi_F(t)\frac{\varkappa}{2\xi^2}\mu_2(t)\alpha_1^{(2)}(t)\big(x_1(t)-
x_2(t)\big)b\big(x_1(t)-x_2(t)\big),\cr
&D_{F,t}^{\alpha}x_2(t)=-\chi_F(t)\frac{\varkappa}{2\xi^2}\mu_1(t)
\alpha_2^{(2)}(t)\big(x_1(t)-x_2(t)\big)b\big(x_1(t)-x_2(t)\big),\label{eesexam1a}\\
&D^{\alpha}_{F,t}(t)\,\alpha_s^{(2)}=2\varepsilon \chi_{F}(t).
\label{eesexam1c}
\end{align}
Equation \eqref{eesexam1c} yields
\begin{equation}
\alpha_s^{(2)}(t)=2\varepsilon S_F^{\alpha}(t)+\alpha_s^{(2)}(0),
\label{eesexamreshc}
\end{equation}
where the initial value $\alpha_s^{(2)}(0)$ is one of the arbitrary constants included in the set $\mathbf{C}$ of integration constants for the system \eqref{eesexam1b}---\eqref{eesexam1c}.
The remaining equations, \eqref{eesexam1a} and \eqref{eesexam1b}, will be analyzed numerically.
\par
Using the conditions \eqref{cauch-1cc}, the arbitrary constants  $\mathbf{C}=\mathbf{C}[\vec{\varphi}]$ in the general solution \eqref{fees3} of the fractal Einstein--Ehrenfest system (Eqs. \eqref{eesexam1a}--\eqref{eesexam1c} and \eqref{eesexamreshc}) can be written in terms of the initial conditions
\begin{align}
&\mu_s(0)=N_s\sigma_s\sqrt{2\pi},\cr
&x_1(0)=-1, x_2(0)=1,\cr
&\alpha_s^{(2)}(0)=\varepsilon\sigma_s^2.
\label{eesexam0}
\end{align}

To illustrate the solutions given by equation \eqref{reshexamp} with initial conditions specified in \eqref{exam1}, and described by the expressions in \eqref{ex-fn-1} --  \eqref{exam2c} and \eqref{eesexam0}, we use the following values for the model parameters:

\begin{align}
&\varepsilon=0.02, \varkappa=1, \xi=2, a=0.5,  b_0=1, N_1=1, N_2=2, \cr
&x_1(0)=-1, x_2(0)=1, \sigma_1=1, \sigma_2=1.5.
	\label{parametr}
\end{align}
To study the behavior of asymptotic solutions with a fractional derivative, we introduce the consideration of moments and asymptotic solutions with an explicit dependence on the parameter $\alpha$, setting $\mu_s=\mu_s(\alpha,t)$, $x_s(\alpha,t)$, $\alpha_s^{(2)}(\alpha,t)$, $u=u(\alpha,x,t)$.
The case $\alpha=1$ corresponds to analytical solutions with the standard first-order time derivative.
To illustrate the influence of fractality on the behavior of quasiparticles, we considered the following values of the parameter $\alpha$: $\log 2/\log4 = 0.5$, $\log2/\log3 \approx 0.63$, $\log2/\log2.5 \approx 0.75$, $\log2/\log2.2 \approx 0.88$, $\log2/\log2.07 \approx 0.95$.
\par
In this example, the corresponding 5th generation prefractals were taken as  approximate Cantor sets.
 The integral staircase function $S_F^{\alpha}(t)$ was constructed using \eqref{cgmss-1}--\eqref{stcf-1} with time-partitions $\delta=10^{-5}$.
\begin{figure}[h]
\centering\includegraphics[scale=0.5]{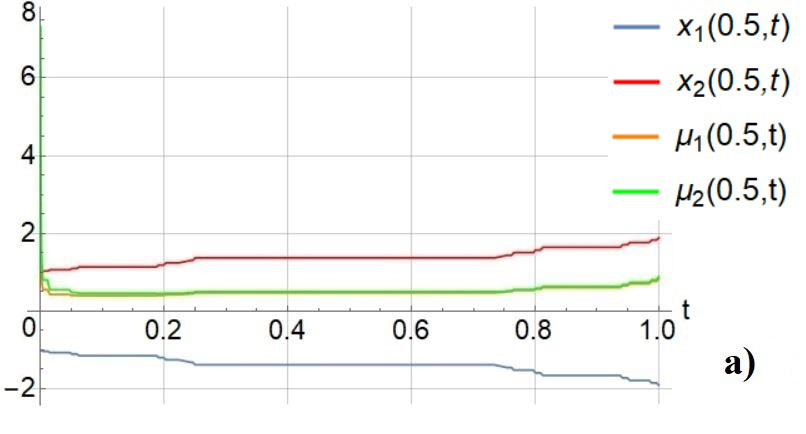}\includegraphics[scale=0.5]{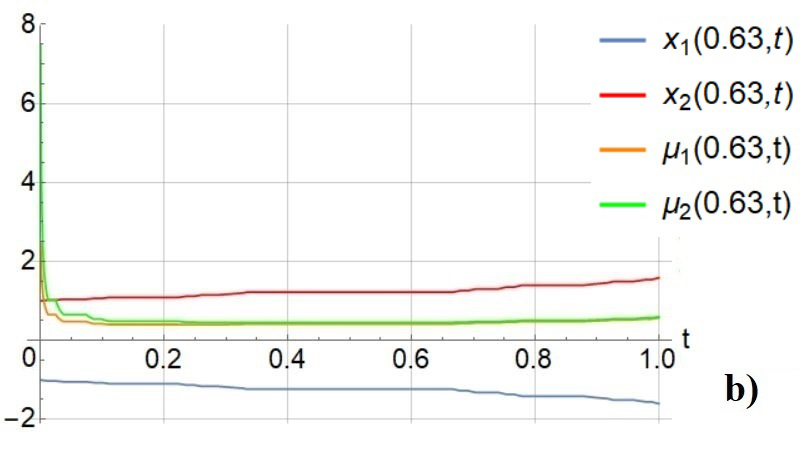}
\includegraphics[scale=0.5]{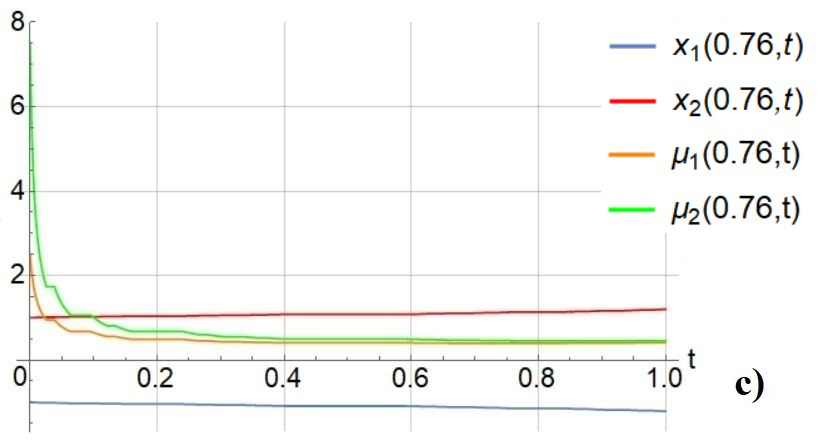}\includegraphics[scale=0.5]{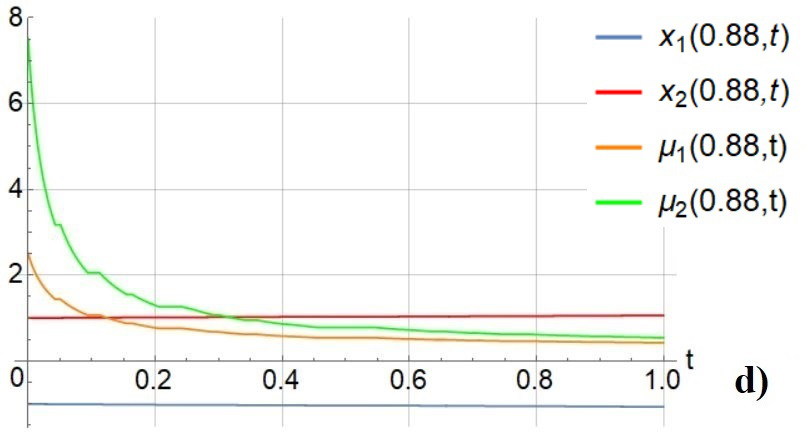}
\end{figure}
\begin{figure}[h]
\centering\includegraphics[scale=0.5]{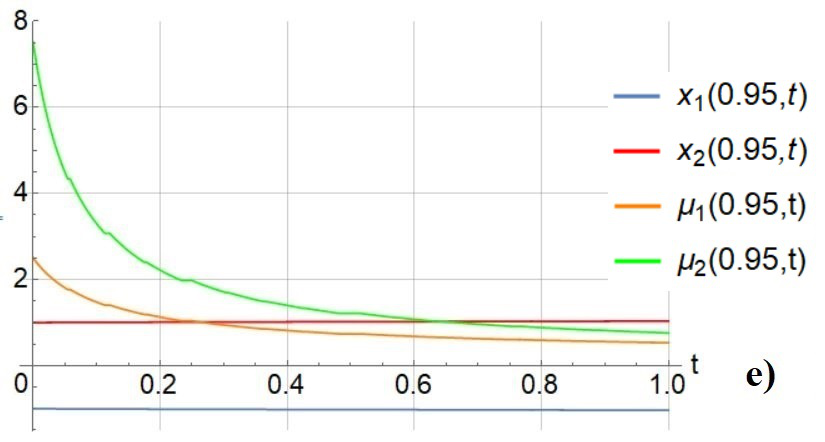}\includegraphics[scale=0.5]{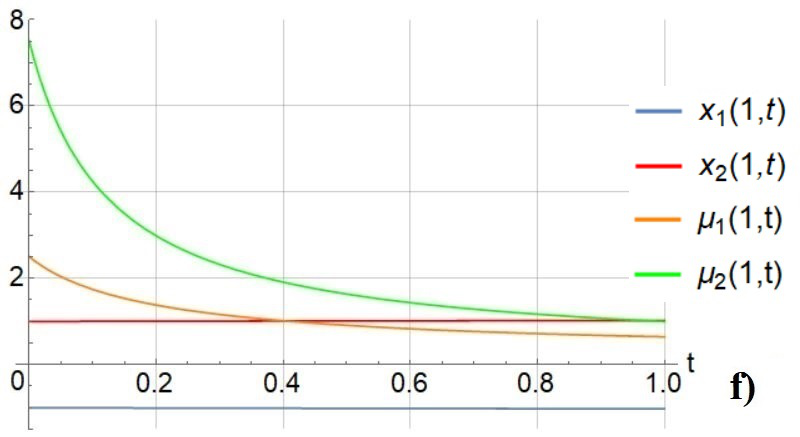}
\caption{\centering{ The zeroth- and  first-order moments,
 $\mu_1(\alpha,t), \mu_2(\alpha,t)$, and $x_1(\alpha,t), x_2(\alpha,t)$, respectively, for two quasiparticles with the parameter $\alpha=0.5$, $0.63$, $0.75$, $0.88$, $0.95$, $1$, are given for  $t \in [0,1]$ and $F=C^{\alpha} \subset [0,1]$.}}\label{pic1}
\end{figure}

Fig. \ref{pic1} shows that as the parameter $\alpha$ decreases, the evolution of the zeroth- and first-order moments accelerates: the zeroth-order moments $\mu_s(\alpha, t)$ reach quasi-stationarity faster as  $t$  increases, and the trajectories $x_s(\alpha, t)$ of quasiparticles
 also diverge more quickly compared to the analytical solutions for moments with the ordinary time derivative, $\mu_s(1,t)$, and $ x_s(1,t)$,   as shown in Fig. \ref{pic1}, panel f).
\begin{figure}[h]
\centering\includegraphics[scale=0.55]{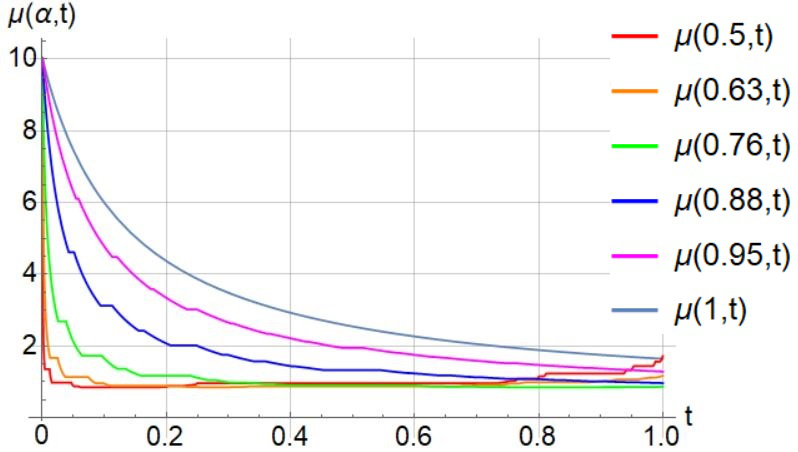}
\caption{\centering{The zeroth-order moments $\mu(\alpha,t)=\mu_1(\alpha,t)+\mu_2(\alpha,t)$ for parameter $\alpha=0.5, 0.63, 0.75, 0.88, 0.95, 1$ and $F=C^{\alpha}\subset[0,1], \, t \in [0,1]$.}}\label{pic2}
\end{figure}

Fig.  \ref{pic2} also illustrates that a decrease in the parameter $\alpha$ accelerates the evolution of the zeroth-order moment. Additionally, as $\alpha$ increases, the total zeroth-order moment approaches the value obtained from the ordinary (first-order) time derivative.
\begin{figure}[H]
\centering\includegraphics[scale=0.5]{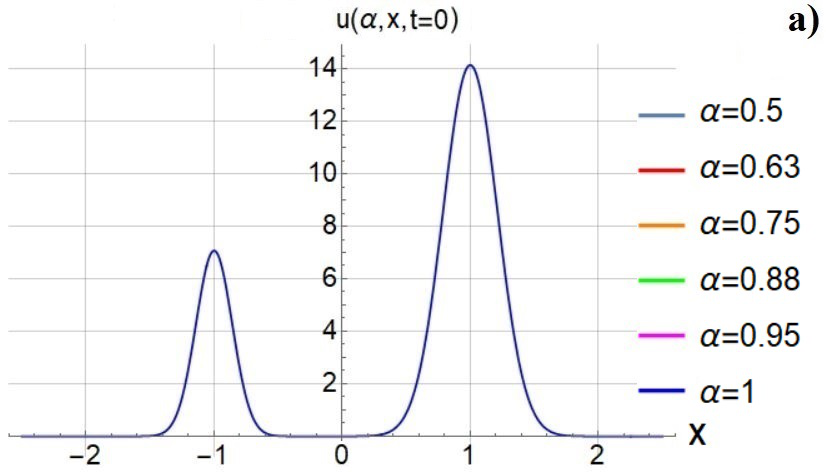}\includegraphics[scale=0.5]{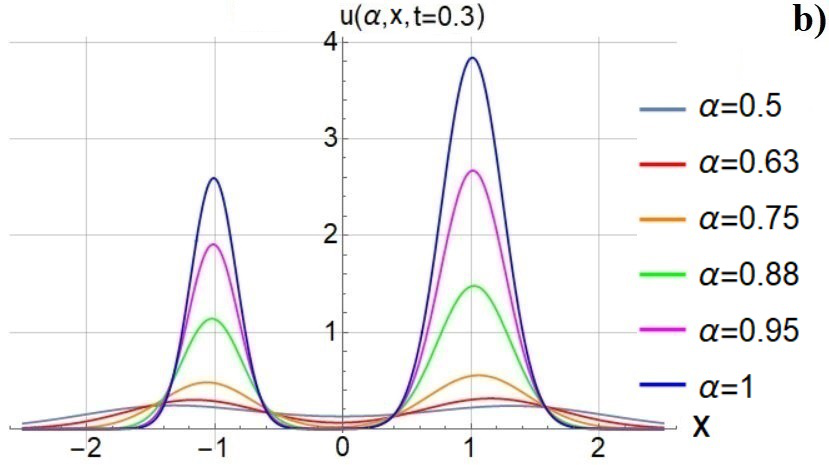}
\includegraphics[scale=0.5]{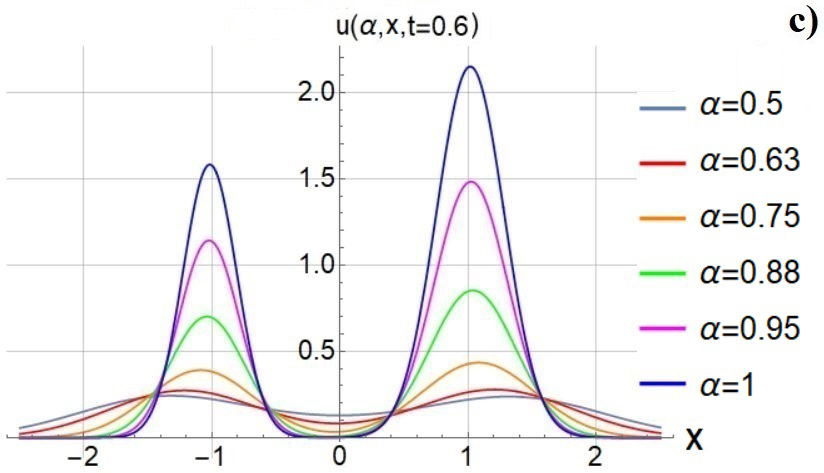}\includegraphics[scale=0.5]{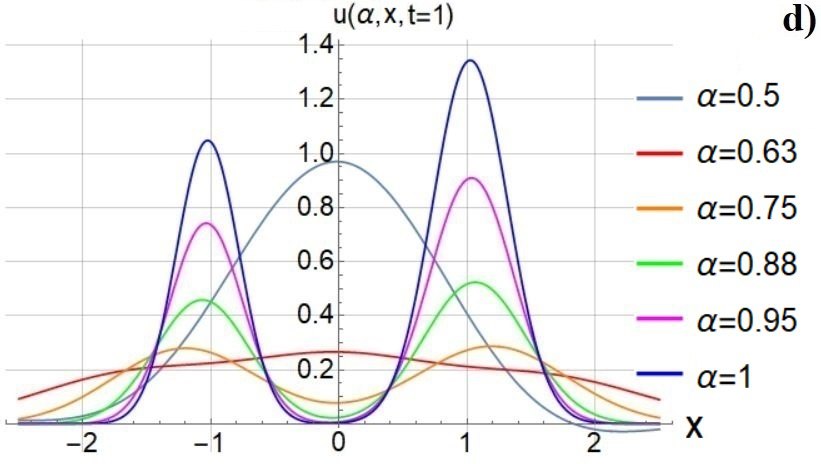}
\caption{\centering{ The asymptotic solutions $u(\alpha,x,t)$ from equation \eqref{sol1a} corresponding to \eqref{solEx} of  equation \eqref{flfkpp1}, are shown for a set of parameters given in \eqref{parametr}, and for $\alpha=0.5$, $0.63$, $0.75$, $0.88$, $0.95$, $1$, at the following time points:  $ a) t=0$, \, $b) t=0.3$, \, $c) t=0.6$, \, $d) t=1$, where $F=C^{\alpha} \subset [0,1]$ and $t \in [0,1]$.}}\label{pic3}
\end{figure}
\par
Fig. \ref{pic3} shows the asymptotic solutions $u(\alpha,x,t)$  from equation \eqref{sol1a} corresponding to \eqref{solEx} of equation \eqref{flfkpp1}, accurate to $O(\varepsilon^{3/2})$, for a set of parameters given in \eqref{parametr}, and for $\alpha=0.5$, $0.63$, $0.75$, $0.88$, $0.95, 1$.
 The solutions are presented at:  $ a) t=0, \, b) t=0.3, \, c) t=0.6, \, d) t=1$, over the Cantor set $F=C^{\alpha} \subset [0,1]$,  where $t \in [0,1]$. During the evolution, the asymmetric initial condition $u(x,0)=\varphi_1(x)+\varphi_2(x)$ (as shown in Fig. \ref{pic3}) evolves toward a symmetric form. As $\alpha$ decreases, the evolution of the asymptotic solutions accelerates. In Fig. \ref{pic3}, one can observe how the processes of diffusion and growth of the total zeroth-order moment (Fig.  \ref{pic2}) destroy the dynamics of quasiparticles for  $\alpha=0.5, 0.63$. In this case, two local peaks merge into a Gaussian-like function, as shown in panel d) of Fig. \ref{pic3}.

\section{Conclusion}
\label{sect-conc}

In this paper, we extend the methodology proposed in \cite{shsin2024} for constructing asymptotic solutions within the weak diffusion approximation to the nonlocal generalized Fisher--KPP equation \eqref{flfkpp1}. This equation incorporates a fractal time derivative of non-integer order $\alpha$, where $0<\alpha\leq 1$. The fractal time derivative is defined within the framework of $F^{\alpha}$-calculus \cite{prgn2009,prgn2011,prsatgn2011},  incorporating the concept of the Maslov  method as discussed in  \cite{maslov1994,beldob1992}.
 \par
The constructed asymptotic solutions exhibit a nontrivial geometric structure, as represented in \eqref{sol1a}. Each solution comprises a finite number of localized distributions (quasiparticles), with each quasiparticle concentrated within a distinct spatial neighborhood that defines its coordinates.
 \par
The example presented in Section \ref{exam-sol1} illustrates the impact of fractality on physical diffusion processes in population dynamics through a solution comprising two quasiparticles. Each quasiparticle initially follows a Gaussian density distribution within the framework of the Fisher--KPP model \eqref{flfkpp1}. The primary objective of this analysis is to examine how the characteristics of these quasiparticles depend on the parameter
$\alpha$  of the fractal time derivative $D^\alpha_{F,t}$. The graphs in Figs.  \ref{pic1}--\ref{pic3} reveal that diffusion governed by the fractal time derivative
$D^\alpha_{F,t}$ with $\alpha <1$  progresses more rapidly than diffusion described by the standard first-order derivative ($\alpha=1$). This acceleration arises from the nature of the underlying time intervals: while ordinary diffusion ($\alpha=1$) evolves continuously over time, fractal diffusion ($\alpha <1$) proceeds discontinuously, with events occurring at discrete points within a fractal Cantor-like time set $F$.
\par
The Einstein--Ehrenfest fractal dynamical system \eqref{fees2a}--\eqref{fees2c} plays a crucial role in constructing asymptotic solutions for the equations \eqref{flsys1a}. This system governs the moments \eqref{mom-11} of the solutions sought for the system \eqref{flsys1a}, which describes quasiparticles and is formulated within function classes of the form \eqref{eq1a}. The FlEES serves as an analogue of the equations of fractal classical mechanics for a system of quasiparticles corresponding to the solution \eqref{sol1a} of equation \eqref{flfkpp1}. The interaction between quasiparticles is determined by the structure of equation \eqref{flfkpp1}.
\par
The mathematical properties of the FlEES are of significant interest in the theory of fractal differential equations, as this system emerges in physical and biological models that describe real-world systems with complex structural properties. Furthermore, the formalism developed here for constructing asymptotic solutions to the fractal equation \eqref{flfkpp1} can be naturally extended to its multidimensional case and applied to other nonlocal generalizations of nonlinear reaction-diffusion dynamics, as demonstrated in studies such as \cite{bbmpv2020, bpv2021, roq2024}, among others.
\par
Note that for the analysis of dynamical systems of the FlEES type, as well as systems of the AFlLE type (described by Eqs. \eqref{fale-1} and \eqref{fale-1b}), it appears promising to employ various ideas and approaches from the symmetry analysis of integer-order partial differential equations (e.g., \cite{obukh2021}, \cite{obukh2022}), methods of integral transformations (e.g., \cite{gscv2025}), various types of approximations (e.g., \cite{awon2025}), and other approaches.
\par
The results presented here provide a foundation for developing approximate methods for the nonlocal Fisher--KPP equation with fractal spatial partial derivatives, a topic of particular interest in both the theory of fractal equations and their applications in biology and physics.

\end{document}